\documentclass[review,3p,times]{elsarticle}
\usepackage{amsmath,hyperref}
\usepackage{lineno}

\usepackage{epstopdf}
\journal{Elsevier}

\usepackage{lineno,hyperref}
\usepackage{graphicx}
\usepackage{dcolumn}
\usepackage{bm}
\usepackage{epsfig}
\usepackage{booktabs}
\usepackage{subfigure}
\usepackage{graphics}
\usepackage{amssymb}
\usepackage{amsmath}
\usepackage{array}
\usepackage{color}
\usepackage{booktabs}
\usepackage{multirow}
\usepackage{caption}
\usepackage{chngpage}
\usepackage{subfigure}
\usepackage{mathrsfs,gensymb,float}

\biboptions{numbers,sort&compress}
\modulolinenumbers[1]

\begin{document}

\captionsetup[figure]{labelfont={bf},name={Fig.},labelsep=period}        

\begin{frontmatter}
	
\title{A new thermal lattice Boltzmann model for liquid-vapor phase change}
\author[mymainaddress,myaddress2]{Lei Wang\corref{mycorrespondingauthor}}
\cortext[mycorrespondingauthor]{Corresponding author}
\ead{wangleir1989@126.com}
\author[mymainaddress,myaddress2]{Jiangxu Huang}
\author[mymainaddress,myaddress2]{Kun He}

\address[mymainaddress]{School of Mathematics and Physics, China University of Geosciences, Wuhan 430074, China}
\address[myaddress2]{Center for Mathematical Sciences, China University of Geosciences, Wuhan 430074, China}

\begin{abstract}
The lattice Boltzmann method is adopted to solve the liquid-vapor phase change problems in this article. By modifying the collision term for the temperature evolution equation, a new thermal lattice Boltzmann model is constructed. As compared with previous studies, the most striking feature of the present approach is that it could avoid the calculations of both the Laplacian term of temperature [$\nabla  \cdot \left( {\kappa \nabla T} \right)$]  and  the gradient term of heat capacitance [$\nabla \left( {\rho {{\rm{c}}_v}} \right)$]. In addition, since the present approach adopts a simple linear equilibrium distribution function, it is possible to use the D2Q5 lattice for the two dimensional cases consided here, making it is more efficiency than previous works in which the lattice is usually limited to the D2Q9. This approach is firstly validated by the problems of droplet evaporation in open space and adroplet evaporation on heated surface, and the numerical results show good agreement with the analytical results and the finite difference method. Then it is used to model nucleate boiling problem, and the relationship between detachment bubble diameter and gravity acceleration obtained with the present approach fits well with the reported works.

\end{abstract}
	
	
\end{frontmatter}

\section{Introduction}
Liquid-vapor phase change processes play a vital role in many industrial applications including nuclear reactor cooling system, power plants, electronic cooling et al. \cite{cho2016nano, wen2018liquid}. For decades, many theoretical and experimental studies have been conducted to reveal the fluid flow and heat transfer during liquid-vapor phase change \cite{widom1970new,citekey2001pool,steinke2004an}. However, due to the various complex phenomena involved in these processes such as interface changes, nonequilibrium effects or other complex dynamic interaction between the phases, the mechanisms of liquid-vapor phase change heat transfer are still not fully comprehended \cite{taylor2009pool,dadhich2019abrief,Kakac2008areview}. With recent advances in computer technology, numerical modelling of such problems has attracted great attention due to its ability to provide the details of flow dynamics during liquid-vapor phase change \cite{liu2015liquid,jamet2001the,son2001anumerical}.

The lattice Boltzmann method (LBM), developted about two decades ago, has gained great success in modelling and simulating of both single-phase and multiphase flows \cite{aidun2010lattice,huang2015multiphase,li2016lattice,huang2021mesoscopic}. Different from traditional computational fluid dynamics methods based on the macroscopic governing equations, the LBM is actually a mesoscopic numerical approach, and its kinetic characteristics bring some distinctive features to this method such as the simple algorithm structure, easy boundary treatment and nature parallelism \cite{kruge2017thelattice,guo2013lattice}. In recent years, the LBM is also adopted by some scholars to simulate liquid-vapor phase change heat transfer such as boiling, evaporation et al., and these existing LB models usually fall into two main categories: (1) the phase-field method \cite{dong2009lattice,sun2013three,safari2013extended}; (2) the pseudopotential method \cite{zhang2003lattice,markus2011simulation,gong2012alattice,li2017improved}. For the phase field method, the vapor-liquid interface is captured by the interface capturing equation like the Cahn-Hilliard equation, while the phase-change process is established by adding a source term to the  Cahn-Hilliard equation, and  the energy equation used in this method is to define the latent heat \cite{sun2013three}. To trigger the liquid-vapor phase change, the phase-field method usually assumes an initial vapor profile in the system \cite{dong2009lattice}, making it cannot be served as an efficient method in modelling bubble nucleation in the boiling heat transfer, while this assumption is not required for the pseudopotential method \cite{li2016lattice}. The key point of pseudopotential method is that the interaction between different phases is mimicked via an attractive or repulsive force among the neighboring fluid particles \cite{shan1993lattice}. As a consequence, the non-ideal gas behavior and phase separation can be realized without using any specific techniques to tarck or capture interfaces \cite{huang2015multiphase,kruge2017thelattice,li2016lattice}. 
Historically, the pioneering work on LBM modelling of liquid-vapor phase change with pseudopotential method may be attributed to Zhang and Cheng in 2003 \cite{zhang2003lattice}. In their work, the  boiling heat transfer is successfully simulated with the proposed method. Then, Hazi and Markus \cite{markus2011simulation} proposed another LB model to simulate the heterogeneous boiling on a horizontal plate . Different from Zhang and Chen's work, the temperature equation appeared in this model is derived from the entropy balance equation and it is coupled with the pseudopotential multiphase LB model through an artificial equqtion of state (EOS). On the basis of Hazi and Markus' work, Gong and Cheng  \cite{gong2012alattice} developted an improved thermal LB model for liquid-vapor phase change heat transfer by employing the Peng-Robinson EOS . Utilizing this model, they successfully simulated the bubble growth and departure in pool boiling. Subsequently, Li et al. \cite{li2017improved} pointed out that the replacement of ${{\nabla  \cdot \left( {\kappa \nabla T} \right)} \mathord{\left/
{\vphantom {{\nabla  \cdot \left( {\kappa \nabla T} \right)} {\left( {\rho {c_v}} \right)}}} \right.\kern-\nulldelimiterspace} {\left( {\rho {c_v}} \right)}}$ with
$\nabla  \cdot \left[ {\left( {{\kappa  \mathord{\left/{\vphantom {\kappa  {\rho {c_v}}}} \right.\kern-\nulldelimiterspace} {\rho {c_v}}}} \right)\nabla T} \right]$ (here, $\kappa$ , ${\rho {c_v}}$ and $T$ are the thermal conductivity, heat capacitance and temperature, respectively) in Gong and Cheng's work is an inappropriate treatment, which will yield considerable errors for the multiphase flows due to the variable density between different phases, and they then proposed another improved thermal LB model for simulation of liuqid-vapor phase change. The same authors also constructed a hybird LB model for liquid-vapor phase change \cite{li2015latticeboil} , in which the velocity field is solved by using the pseudopotential LB approach, while the temperature field is solved by the finite-difference method. Thereafter,  a thermal multiple-relaxation-time (MRT) LB model with nondiagonal matrix based on the two-dimensional nine-velocity (D2Q9) lattice is developted for liquid-vapor phase change by Zhang et al. \cite{zhang2021improved} . Different from Li et al.'s model, the calculation of the Laplacian term of temperature is avoided in this model, and the latent heat of vaporization is also decoupled with the EOS, making it is more flexibility in simulating liquid-vapor phase change. However,  due to the nondiagonal matrix appeared in this approach relies on the lattice model, it can not be extended to the three-dimensional (3D) case directly. More recently,  a three-dimensional thermal LB model is constructed for liquid-vapor phase change by Li et al. \cite{{li2022improvedlattice}}. Different from previous works, this model is possible to use the D3Q7 lattice due to the convection term in the corresponding LB equation is actually treated as a source term,  while it still needs to calculate the gradient term of heat capacitance $\nabla \left( {\rho {c_v}} \right)$.

In this work, we propose a new thermal LB model for liquid-vapor phase change. By modifying the collision term for the temperature equation, the calculations of the Laplacian term of temperature [$\nabla  \cdot \left( {\kappa \nabla T} \right)$]  and  the gradient term of heat capacitance [$\nabla \left( {\rho {{\rm{c}}_v}} \right)$]  are both avoided, resulting it retains the main advantages of original LBM.  Moreover, due to the present thermal model is constructed based on a linear equilibrium distribution, it is possible to use a more simple D2Q5 lattice, and also a D3Q7 lattice when it extended to 3D space. The rest of the paper is organized as follows. In Section 2, a pesudopotential model is briefly introduced. Then, a new thermal LB model is proposed in Section 3. Numerical validation of the proposed model is presented in Section 4. Finally, a brief summary is concluded in Section 5.

\section{Pseudopotential lattice Boltzmann model}
The pseudopotential multiphase model, originally proposed by Shan and Chen \cite{shan1993lattice}, is a particularly popular method  in the LB community for its simplicity and intuitive connection to classical non-ideal gas EOS \cite{yuan2006equations} . To achieve the thermodynamic consistency, various improved pseudopotential model are proposed in the past two decades \cite{li2013latticeimproved,kupershtokh2009on,huangjcp2016}. In this work,  an improved D2Q9 multiple-relaxation-time (MRT) pseudopotential model developted by Li et al. \cite{li2013latticeimproved} is considered, and the evolution equation of the density distribution  ${f_i}$ in this model is written as 
\begin{equation}
{f_i}\left( {{\bf{x}} + {{\bf{c}}_i}\Delta t,t + \Delta t} \right) - {f_i}\left( {{\bf{x}},t} \right) =  - {\left( {{{\bf{M}}^{ - 1}}{\bf{SM}}} \right)_{ij}}\left[ {{f_j}\left( {{\bf{x}},t} \right) - f_j^{(eq)}\left( {{\bf{x}},t} \right)} \right] + \Delta t{{F'}_i}\left( {{\bf{x}},t} \right),\label{eq1}
\end{equation}
in which ${{{\bf{c}}_i}}$ is the discrete velocity at position ${\bf{x}}$ and time $t$. For the D2Q9 model considered here, the discrete velocities are defined as \cite{kruge2017thelattice} 
\begin{equation}
{{\bf{c}}_i} = \left\{ \begin{array}{l}
\left( {0,0} \right),\;\;\;\;\;\;\;\;\;\;\;\;\;\;\;\;\;\;\;\;\;\;\;\;\;\;\;\;\;\;\;\;\;\;\;\;\;\;\;\;\;\;\;\;\;\;\;\;\;\;\;\;\;\;i = 0,\\
\left( {\cos \left[ {\left( {i - 1} \right){\pi  \mathord{\left/
{\vphantom {\pi  2}} \right.
\kern-\nulldelimiterspace} 2}} \right],\sin \left[ {\left( {i - 1} \right){\pi  \mathord{\left/{\vphantom {\pi  2}} \right.
\kern-\nulldelimiterspace} 2}} \right]} \right)c,\;\;\;\;\;\;\;\;\;\;\;\;i = 1 - 4,\\
\left( {\cos \left[ {\left( {2i - 9} \right){\pi  \mathord{\left/
{\vphantom {\pi  4}} \right.
\kern-\nulldelimiterspace} 4}} \right],\sin \left[ {\left( {2i - 9} \right){\pi  \mathord{\left/{\vphantom {\pi  4}} \right.\kern-\nulldelimiterspace} 4}} \right]} \right)\sqrt 2 c,\;\;\;i = 5 - 8,
	\end{array} \right.
\end{equation}
where $c = {{\Delta x} \mathord{\left/ {\vphantom {{\Delta x} {\Delta t}}} \right.
\kern-\nulldelimiterspace} {\Delta t}}$ is the lattice speed with $\Delta x$ and $\Delta t$ denoting the lattice spacing and time step (both of them are set to be 1 in the present work). ${f_i^{(eq)}\left( {{\bf{x}},t} \right)}$ is the equilibrium distribution given by \cite{chen1998lattice}

\begin{equation}
f_i^{(eq)}\left( {{\bf{x}},t} \right) = {\omega _i}\rho \left[ {1 + \frac{{{{\bf{c}}_i} \cdot {\bf{u}}}}{{c_s^2}} + \frac{{{{\left( {{{\bf{c}}_i} \cdot {\bf{u}}} \right)}^2}}}{{2c_s^4}} - \frac{{{{\left| {\bf{u}} \right|}^2}}}{{2c_s^2}}} \right],
\end{equation} 
where  ${\omega _i}$ is the the weight coefficient given by  ${\omega _0} = {4 \mathord{\left/{\vphantom {4 9}} \right.\kern-\nulldelimiterspace} 9},{\omega _{1 - 4}} = {1 \mathord{\left/{\vphantom {1 9}} \right.\kern-\nulldelimiterspace} 9},{\omega _{5 - 8}} = {1 \mathord{\left/{\vphantom {1 {36}}} \right.\kern-\nulldelimiterspace} {36}}$ , ${\bf{u}} = \left( {{u_x},{u_y}} \right)$ is the velocity and ${c_s} = {c \mathord{\left/{\vphantom {c {\sqrt 3 }}} \right.\kern-\nulldelimiterspace} {\sqrt 3 }}$ is the sound speed.  ${\bf{M}}$ is the transformation matrix given by \cite{lallemand2000theory}
\begin{equation}
	\left[ {\begin{array}{*{20}{c}}
			1&1&1&1&1&1&1&1&1\\
			{ - 4}&{ - 1}&{ - 1}&{ - 1}&{ - 1}&2&2&2&2\\
			4&{ - 2}&{ - 2}&{ - 2}&1&1&1&1&1\\
			0&1&0&{ - 1}&0&1&{ - 1}&{ - 1}&1\\
			0&{ - 2}&0&2&0&1&{ - 1}&{ - 1}&1\\
			0&0&1&0&{ - 1}&1&1&{ - 1}&{ - 1}\\
			0&0&{ - 2}&0&2&1&1&{ - 1}&{ - 1}\\
			0&1&{ - 1}&1&{ - 1}&0&0&0&0\\
			0&0&0&0&0&1&{ - 1}&1&{ - 1}
	\end{array}} \right].   
\end{equation}
Note that when Eq. (\ref{eq1}) is multiplied by the transformation matrix ${\bf{M}}$,   the original collision step implemented in velocity space will turn to execute in the momentum space, 
\begin{equation}
{{\bf{m}}^*}\left( {{\bf{x}},t} \right) = {\bf{m}}\left( {{\bf{x}},t} \right) - {\bf{S}}\left[ {{\bf{m}}\left( {{\bf{x}},t} \right) - {{\bf{m}}^{eq}}\left( {{\bf{x}},t} \right)} \right] + \Delta t\left( {{\bf{{\rm I}}} - \frac{{\bf{S}}}{2}} \right){\bf{F}}\left( {{\bf{x}},t} \right),
\end{equation} 
while the streaming step is still implemented in the velocity space
\begin{equation}
{f_i}\left( {{\bf{x}} + {{\bf{c}}_i}\Delta t,t + \Delta t} \right) = f_i^*\left( {{\bf{x}},t} \right).
\end{equation}
Here,  ${\bf{m}}\left( {{\bf{x}},t} \right) = {\bf{Mf}}$ is the rescaled moment with ${\bf{f}}\left( {{\bf{x}},t} \right) = {\left[ {{f_0}\left( {{\bf{x}},t} \right), \cdots ,{f_8}\left( {{\bf{x}},t} \right)} \right]^{\rm T}}$, $f_i^*\left( {{\bf{x}},t} \right) = {{\bf{M}}^{ - 1}}{{\bf{m}}^*}\left( {{\bf{x}},t} \right)$ is the post-collision distribution function, ${\bf{{\rm I}}}$ is the unit matrix, ${\bf{S}} = {\rm{diag}}\left( {{s_0},{s_e},{s_\varepsilon },{s_j},{s_q},{s_j},{s_q},{s_p},{s_p}} \right)$ is the diagonal relaxation matrix. ${{{\bf{m}}^{eq}}\left( {{\bf{x}},t} \right)}$ is the equilibrium function in the moment space given by 
\begin{equation}
{{\bf{m}}^{eq}}\left( {{\bf{x}},t} \right) = \rho {\left[ {1, - 2 + 3{{\left| {\bf{u}} \right|}^2},1 - 3{{\left| {\bf{u}} \right|}^2},{u_x}, - {u_x},{u_y}, - {u_y},u_x^2 - u_y^2,{u_x}{u_y}} \right]^{\rm T}}.
\end{equation}
In addition,  ${\bf{F}}\left( {{\bf{x}},t} \right)$ is the forcing term in the moment space satisfying $\left( {{\bf{{\rm I}}} - 0.5{\bf{S}}} \right){\bf{F}}\left( {{\bf{x}},t} \right) = {\bf{MF'}}$. Following the work of Li et al.\cite{li2013latticeimproved}, it can be written as
\begin{equation}
	{\bf{F}} = \left[ {\begin{array}{*{20}{c}}
			0\\
			{6{\bf{u}} \cdot {{\bf{F}}_n} + \frac{{\sigma {{\left| {{{\bf{F}}_m}} \right|}^2}}}{{{\psi ^2}\Delta t\left( {{s_e}^{ - 1} - 0.5} \right)}}}\\
			{ - 6{\bf{u}} \cdot {{\bf{F}}_n} - \frac{{\sigma {{\left| {{{\bf{F}}_m}} \right|}^2}}}{{{\psi ^2}\Delta t\left( {{s_\varepsilon }^{ - 1} - 0.5} \right)}}}\\
			{{F_{n,x}}}\\
			{ - {F_{n,x}}}\\
			{{F_{n,y}}}\\
			{ - {F_{n,y}}}\\
			{2\left( {{u_x}{F_{n,x}} - {u_y}{F_{n,y}}} \right)}\\
			{{u_x}{F_{n,y}} + {u_y}{F_{n,x}}}
	\end{array}} \right]
\end{equation}
where $\sigma$ is a tunable parameter used to achieve thermodynamic consistency,   ${{\bf{F}}_n} = \left( {{F_{n,x}},{F_{n,y}}} \right)$ is the total force, and it is defined as ${{\bf{F}}_n} = {{\bf{F}}_m} + {{\bf{F}}_b}$, with ${{\bf{F}}_b}$ and ${{\bf{F}}_m}$ representing the external force and the  intermolecular interaction force, respectively. For the D2Q9 lattice considered here, the interaction force is given by \cite{shan2008pressure}
\begin{equation}
	{{\bf{F}}_m} =  - G\psi \left( {\bf{x}} \right)\sum\limits_{i = 1}^8 {\varpi \left( {{{\left| {{{\bf{c}}_i}\Delta t} \right|}^2}} \right)\psi \left( {{\bf{x}} + {{\bf{c}}_i}\Delta t} \right)} {{\bf{c}}_i}\Delta t
\end{equation} 
where ${\varpi \left( {{{\left| {{{\bf{c}}_i}\Delta t} \right|}^2}} \right)}$ are the weights, which are given as $\varpi \left( {\Delta {x^2}} \right) = {1 \mathord{\left/{\vphantom {1 3}} \right.
\kern-\nulldelimiterspace} 3},\varpi \left( {2\Delta {x^2}} \right) = {1 \mathord{\left/{\vphantom {1 {12}}} \right.\kern-\nulldelimiterspace} {12}}$, $G$ is a parameter that controls the strength of the inter-particle force, whose value is usually set to be -1 in many simulations, $\psi \left( {\bf{x}} \right)$ is the mean-field potential, which is calculated from $\psi \left( {\bf{x}} \right) = \sqrt {{{2\left( {{p_{EOS}} - \rho c_s^2} \right)} \mathord{\left/
{\vphantom {{2\left( {{p_{EOS}} - \rho c_s^2} \right)} {\left( {G\Delta {x^2}} \right)}}} \right.\kern-\nulldelimiterspace} {\left( {G\Delta {x^2}} \right)}}} $ with ${{p_{EOS}}}$ is a non-ideal EOS \cite{yuan2006equations}. In the present work, the Peng-Robinson (P-R) EOS is used \cite{huang2015multiphase}, which is given as 
\begin{equation}
	{p_{EOS}} = \frac{{\rho RT}}{{1 - b\rho }} - \frac{{a\varphi \left( T \right){\rho ^2}}}{{1 + 2b\rho  - {b^2}{\rho ^2}}},
\end{equation} 
where  $R$ is the gas constant, $a = {{0.45724{R^2}T_c^2} \mathord{\left/{\vphantom {{0.45724{R^2}T_c^2} {{p_c}}}} \right. \kern-\nulldelimiterspace} {{p_c}}}$ and $b = {{0.0778R{T_c}} \mathord{\left/
{\vphantom {{0.0778R{T_c}} {{p_c}}}} \right.
\kern-\nulldelimiterspace} {{p_c}}}$, with ${T_c}$ and ${p_c}$ representing the critical temperature and the critical pressure, respectively. $\varphi \left( T \right) = {\left[ {1 + \left( {0.37464 + 1.54226\bar \omega  - 0.26992{{\bar \omega }^2}} \right)\left( {1 - \sqrt {{T \mathord{\left/{\vphantom {T {{T_c}}}} \right.\kern-\nulldelimiterspace} {{T_c}}}} } \right)} \right]^2}$, in which  ${\bar \omega }=0.344$ is the acentric factor, $T$ is the temperature, and it is calculated from the energy equation in the real simulations.  For the other parameters, we choose  $a = {3 \mathord{\left/{\vphantom {3 {49}}} \right.\kern-\nulldelimiterspace} {49}},b = {2 \mathord{\left/{\vphantom {2 {21}}} \right.\kern-\nulldelimiterspace} {21}}$ and $R = 1$, which have been widely used in previous works. Finally, the macroscopic density $\rho$ and velocity $\bf{u}$ in pseudopotential model are calculated by
\begin{equation}
	\rho  = \sum\limits_{i = 0}^8 {{f_i}} ,\;\;\;\;\;\;\;\;\;\;\;\;\rho {\bf{u}} = \sum\limits_{i = 0}^8 {{{\bf{c}}_i}{f_i}}  + \frac{{\Delta t}}{2}{{\bf{F}}_n}.
\end{equation} 

\section{Numerical formulations}
\subsection{Reexamination of previous lattice Boltzmann models }
\label{set31}

Before proceeding further, the governing equation for liquid-vapor phase change is first revisited.  According to the work of Anderson et al. \cite{anderson1998diffuse},   by neglecting the effect of viscous heat dissipation, the energy equation for liquid-vapor phase change can be expressed as (also called local balance law for entropy)
\begin{equation}
	\rho T\frac{{Ds}}{{Dt}} = \nabla  \cdot \left( {\lambda \nabla T} \right),
\end{equation}
where ${{D\left(  \cdot  \right)} \mathord{\left/{\vphantom {{D\left(  \cdot  \right)} {Dt}}} \right. \kern-\nulldelimiterspace} {Dt}} = {\partial _t}\left(  \cdot  \right) + {\bf{u}} \cdot \nabla \left(  \cdot  \right)$ is the material derivative, $s$ is the entropy and $\lambda $ is the thermal conductivity. 
To simplify the above equation, the following thermodynamic relation is considered \cite{bird2001transport}
\begin{equation}
ds = \frac{{{c_v}}}{T}dT + {\left( {\frac{{\partial {p_{EOS}}}}{{\partial T}}} \right)_\rho }d\left( {\frac{1}{\rho }} \right).
\label{eq13}
\end{equation}
in which  $c_v$ is the specific heat at constant volume. According to Eq. (\ref{eq13}) and note that the continuity equation ${{D\rho } \mathord{\left/{\vphantom {{D\rho } {Dt}}} \right.\kern-\nulldelimiterspace} {Dt}} =  - \rho \nabla  \cdot {\bf{u}}$  , the temperature equation for liquid-vapor phase change can be written as  

\begin{equation}
\rho {c_v}\frac{{\partial T}}{{\partial t}} + \rho {c_v}{\bf{u}} \cdot \nabla T = \nabla  \cdot \left( {\lambda \nabla T} \right) - T{\left( {\frac{{\partial {p_{EOS}}}}{{\partial T}}} \right)_\rho }\nabla  \cdot {\bf{u}}.
\label{eq14}
\end{equation}

In order to solve the above equation within the framework of LBM, various models have been proposed in the previous works \cite{gong2012alattice,li2017improved,zhang2021improved,li2022improvedlattice}, however, the evolution equations appeared in these existing works are rather heuristically. In fact, to match the corrresponding thermal LB models, these models are almost constructed based on the following variant temperature equation \cite{gong2012alattice,li2017improved,zhang2021improved,li2022improvedlattice} 
\begin{equation}
\frac{{\partial T}}{{\partial t}} + \nabla  \cdot \left( {{\bf{u}}T} \right) = \nabla  \cdot \left( {\eta \nabla T} \right) + R,
\label{eq15}
\end{equation}
where $\eta$ is the thermal diffusivity or an artificial parameter depending on the expression of the  source term of $R$.  Since Eq. (\ref{eq15}) is a standard convection-diffusion equation with a source term, there are many universal models can be adopted for solving this equation in the LB community. Following this idea, Gong and Cheng \cite{gong2012alattice} proposed a LB model for liquid-vapor phase change where  
\begin{equation}
\eta  = \frac{\lambda }{{\rho {c_v}}},\;\;\;\;\;R = T\left[ {1 - \frac{1}{{\rho {c_v}}}{{\left( {\frac{{\partial {p_{EOS}}}}{{\partial T}}} \right)}_\rho }} \right]\nabla  \cdot {\bf{u}}.
\label{eq16}
\end{equation}
Comparing Eq. (\ref{eq16}) with Eq. (\ref{eq14}), one can found that the term of ${{\nabla  \cdot \left( {\lambda \nabla T} \right)} \mathord{\left/
{\vphantom {{\nabla  \cdot \left( {\lambda \nabla T} \right)} {\left( {\rho {c_v}} \right)}}} \right.	\kern-\nulldelimiterspace} {\left( {\rho {c_v}} \right)}}$ is replaced by $\nabla  \cdot \left\{ {\left[ {{\lambda  \mathord{\left/
{\vphantom {\lambda  {\left( {\rho {c_v}} \right)}}} \right.
\kern-\nulldelimiterspace} {\left( {\rho {c_v}} \right)}}} \right]\nabla T} \right\}$ in Gong and Cheng's model. Although this treatment is acceptable  for the single-phase flow in the incompressible limit, it is not correct for the multiphase flows especially for the liquid-vapor interface, where the density varies significantly. In this setting, Li et al. \cite{li2017improved} proposed an improved LBM, and in their model,   
\begin{equation}
\eta  = k,\;\;\;\;\;R = \frac{1}{{\rho {c_v}}}\nabla  \cdot \left( {\lambda \nabla T} \right) - \nabla  \cdot \left( {k\nabla T} \right) + T\left[ {1 - \frac{1}{{\rho {c_v}}}{{\left( {\frac{{\partial {p_{EOS}}}}{{\partial T}}} \right)}_\rho }} \right]\nabla  \cdot {\bf{u}},	
\end{equation}
where $k$ is an artificial parameter. For this model, due to the source term $R$ is related to the Laplacian/gradient term of temperature, one must calculate it with the help of finite-difference scheme, making it can not hold the advantage of the LBM very well. In addition, Zhang et al. \cite{zhang2021improved} also proposed another improved thermal LBM for liquid-vapor phase change where 
\begin{equation}
\eta  = \frac{\lambda }{{\rho {c_v}}},\;\;\;\;\;R = \frac{{\lambda \nabla T \cdot \nabla \left( {\rho {c_v}} \right)}}{{{{\left( {\rho {c_v}} \right)}^2}}} + \left( {T - \frac{{\rho H}}{{{c_v}}} - \frac{{{p_{EOS}}}}{{\rho {c_v}}}} \right)\nabla  \cdot {\bf{u}},
\end{equation}
in which $H$ is a parameter related to the latent heat of vaporization. In contrast to Li et al.'s model, this model is not necessary to calculate the Laplacian of temperature, and the latent heat is decoupled with the EOS, but it still needs to calculate the gradient term of ${\nabla \left( {\rho {c_v}} \right)}$ with the finite-difference scheme, which will bring some numerical errors into the system, and this phenomenon may be more remarkable around the liquid-vapor interface. On the other hand, since the calculation of temperature in Zhang et al.'s model is related to the source term $R$, causing the corresponding expression to be implicit, and an iteration procedure is needed theoretically. Apart from the above models, Li et al. \cite{li2022improvedlattice} recently also proposed a 3D thermal LBM for phase change heat transfer in which 
\begin{equation}
\eta  = \frac{\lambda }{{\rho {c_v}}},\;\;\;\;\;R = \frac{{\lambda \nabla T \cdot \nabla \left( {\rho {c_v}} \right)}}{{{{\left( {\rho {c_v}} \right)}^2}}} - {\bf{u}} \cdot \nabla T - \frac{T}{{\rho {c_v}}}{\left( {\frac{{\partial {p_{EOS}}}}{{\partial T}}} \right)_\rho }\nabla  \cdot {\bf{u}} + \nabla  \cdot \left( {{\bf{u}}T} \right).
\label{eq19}
\end{equation}
Substituting Eq. (\ref{eq19}) into Eq. (\ref{eq15}), one can find that the convection term in the temperature equation Eq. (\ref{eq14}) is actually treated as a source term in this model, however, the calculation of  ${\nabla \left( {\rho {c_v}} \right)}$ still exists. 
\subsection{New thermal lattice Boltzmann model}
In this section, we propose a new thermal LB model for liquid-vapor phase change. To this end,  the temperature equation is changed to   
\begin{equation}
\rho {c_v}\frac{{\partial T}}{{\partial t}} = \nabla  \cdot \left( {\lambda \nabla T} \right) - \left[ {\rho {c_v}{\bf{u}} \cdot \nabla T + T{{\left( {\frac{{\partial {p_{EOS}}}}{{\partial T}}} \right)}_\rho }\nabla  \cdot {\bf{u}}} \right].
\label{eq20}
\end{equation}
Apparently, the above equation can be viewed as a pure diffusion equation $\rho {c_v}{\partial _t}T = \nabla  \cdot \left( {\lambda \nabla T} \right) + Q$ with a corresponding source term $Q =  - \left[ {\rho {c_v}{\bf{u}} \cdot \nabla T + T{{\left( {{{\partial {p_{EOS}}} \mathord{\left/{\vphantom {{\partial {p_{EOS}}} {\partial T}}} \right.\kern-\nulldelimiterspace} {\partial T}}} \right)}_\rho }\nabla  \cdot {\bf{u}}} \right]$. Although it is not difficult to develop a LBM for solving pure diffusion equation, how to incorporate $\rho {c_v}$ in front of ${{\partial T} \mathord{\left/{\vphantom {{\partial T} {\partial t}}} \right.
\kern-\nulldelimiterspace} {\partial t}}$ into the temperature evolution equation is a problem that must be addressed. To have a better understand on the proposed LBM, we first present the model by using the simplest Bhatnagar-GrossKrook (BGK) operator, and then extend it to the  multiple-relaxationtime (MRT) model, which is a generalized model  and has distinct advantages over the BGK model in terms of stability and accuracy. 
\subsubsection{BGK model for liquid-vapor phase change}
The lattice BGK equation for the temperature distribution function ${g_i}$ is expressed as 
\begin{equation}
	\rho c_v{g_i}\left( {{\bf{x}} + {{\bf{c}}_i}\Delta t,t + \Delta t} \right) -
	{g_i}\left( {{\bf{x}},t} \right)  =(\rho c_v-1){g_i}\left( {{\bf{x}} + {{\bf{c}}_i}\Delta t,t} \right)
	- \frac{1}{\tau_g}\left[
	{{g_i}\left( {{\bf{x}},t} \right) - g_i^{(eq)}\left( {{\bf{x}},t}
		\right)} \right]+\Delta t {G_i} + \Delta t {S_i}  ,
\end{equation} 
where $\tau_g$ is the relaxation time, $ g_i^{(eq)}$ is the local equilibrium distribution function defined as 
\begin{equation}
	g_i^{(eq)} = {{\hat w}_i}T,
	\label{eq22}
\end{equation} 
$G_i$ is the source term given by 
\begin{equation}
{G_i} =  - {{\hat w}_i}\left[ {\rho {c_v}{\bf{u}} \cdot \nabla T + T{{\left( {{\partial _T}{p_{EOS}}} \right)}_\rho }\nabla  \cdot {\bf{u}}} \right],
	\label{eq23}
\end{equation}
and the correction term $S_i$ is chosen as 
\begin{equation}
	S_i={{\hat w}_i} \rho c_v\frac{\Delta t}{2} \partial_t^2T.
		\label{eq24}
\end{equation}
The temperature $T$ is calculated from
\begin{equation}
T = \sum\limits_i {{g_i}}.
\end{equation} 

Since the equilibrium distribution function is a linear form, it is possible to use a more simple D2Q5 lattice model, in which the discrete velocity set is expressed as 
\begin{equation}
	{{\bf{c}}_i} = \left\{ \begin{array}{l}
		\left( {0,0} \right),\;\;\;\;\;\;\;\;\;\;\;\;\;\;\;\;\;\;\;\;\;\;\;\;\;\;\;\;\;\;\;\;\;\;\;\;\;\;\;\;\;\;\;\;\;\;\;\;\;\;\;\;\;\;i = 0\\
		\left( {\cos \left[ {\left( {i - 1} \right){\pi  \mathord{\left/
						{\vphantom {\pi  2}} \right.
						\kern-\nulldelimiterspace} 2}} \right],\sin \left[ {\left( {i - 1} \right){\pi  \mathord{\left/
						{\vphantom {\pi  2}} \right.
						\kern-\nulldelimiterspace} 2}} \right]} \right)c,\;\;\;\;\;\;\;\;\;\;\;\;i = 1 - 4
	\end{array} \right.
\end{equation}   
and the corresponding weight coefficient in such a case can be defined as ${{\hat w}_{i = 0}} = 1 - \dot w $, ${{\hat w}_{i = 1 - 4}} = {{\dot w} \mathord{\left/
		{\vphantom {{\dot w} 4}} \right.
		\kern-\nulldelimiterspace} 4}$ (${\hat w}$ is a parameter satisfying $\dot w \in \left( {0,1} \right)$)  with the sound speed given by $\hat c_s^2 = {{\dot w} \mathord{\left/
		{\vphantom {{\dot w} 2}} \right.
		\kern-\nulldelimiterspace} 2}$ \cite{he2019latticeBo}.

To recover the macroscopic temperature equation from the lattice BGK eqaution, in what follows we will perform a multi-scale analysis of the present model. To this end, the distribution function $g_i$ and the time and space derivaties are first expanded as \cite{guo2013lattice}
\begin{equation}
{g_i} = g_i^{\left( 0 \right)} + \epsilon g_i^{\left( 1 \right)} + {\epsilon ^2}g_i^{\left( 2 \right)},\;\;\;\;\;{\partial _t} = \epsilon {\partial _{{t_1}}} + {\epsilon ^2}{\partial _{{t_2}}},\;\;\;\;\;\nabla  = \epsilon {\nabla _1},
\label{eq27}
\end{equation}
where $t_1$ is the fast convective scale, $t_2$ is the slow diffusive scale, and $\epsilon$ is a small parameter which is propportional to Knudsen number in the classical kinetic theory for fluid flows. 

By Taylor expanding the lattice BGK equation yields
\begin{equation}
\rho c_v\left[g_i+ \Delta t D_ig_i+\frac{\Delta t^2}{2}D_i^2g_i    \right] -
g_i  +(1-\rho c_v)\left[g_i+\Delta td_i g_i +\frac{\Delta t^2}{2}d_i^2g_i  \right]
=- \frac{1}{\tau_g}\left[{g_i - g_i^{(eq)}} \right]   +\Delta t {G_i}+\Delta t {S_i},
\label{eq28}
\end{equation}
where ${D_i} = {\partial _t} + {d_i}$ with ${d_i} = {{\bf{c}}_i} \cdot \nabla$. Then substituting the expansions Eq. (\ref{eq27}) into  Eq. (\ref{eq28}), and equating the coefficients of each order of $\epsilon$, we have 
\begin{equation}
{\rm O}(\epsilon ^0):\;\; g_i^{(0)}=g_i^{(eq)}
\label{eq29}
\end{equation}

\begin{equation}
{\rm O}(\epsilon ^1):\;\; \rho c_v\partial_ {t_1}g_i^{(0)}+d_{1i}g_i^{(0)}=- \frac{1}{{\tau_g} \Delta t}g_i^{(1)} + {{\hat w}_i Q^{(1)}} 	
\label{eq30}
\end{equation}

\begin{equation}       
{\rm O}(\epsilon ^2): \;\;\rho c_vD_{1i}g_i^{(1)}+\rho c_v\partial_ {t_2}g_i^{(0)}+\rho c_v \frac{\Delta t}{2}D_{1i}^2g_i^{(0)}+(1-\rho c_v)d_{1i}g_i^{(1)}+(1-\rho c_v)\frac{\Delta t}{2} d_{1i}^2g_i^{(0)}=- \frac{1}{{\tau_g} \Delta t}g_i^{(2)}+{\hat w}_i \rho c_v\frac{\Delta t}{2} \left(\partial t_1\right)^2T 
\label{eq31}
\end{equation}
where ${D_{{1i}}} = {\partial _{{t_1}}} + {d_{{1i}}} = {\partial _{{t_1}}} + {{\bf{c}}_i} \cdot {\nabla _1}$. If we expand ${D_{{1i}}}$ and $D_{{1i}}^2$ in Eq. (\ref{eq31}), then the corresponding equation can be rewritten as  
\begin{equation}
\rho c_v\partial_{t_1}g_i^{(1)}+  d_{1i}g_i^{(1)} +\rho c_v\partial_ {t_2}g_i^{(0)}+\rho c_v \frac{\Delta t}{2}\partial_{t_1}^2g_i^{(0)}+\rho c_v \frac{\Delta t}{2}\left(2\partial_{t_1}d_{1i}\right)g_i^{(0)}+ \frac{\Delta t}{2}d_{1i}^2g_i^{(0)}=- \frac{1}{{\tau_g} \Delta t}g_i^{(2)}+{\hat w}_i \rho c_v\frac{\Delta t}{2} \left(\partial t_1\right)^2T 
\label{eq32}
\end{equation}
From Eq. (\ref{eq22})-Eq. (\ref{eq24}) and Eq. (\ref{eq29}), one can obtain that 
\begin{equation}
\sum_{i}g_i^{(0)}=T, \quad \sum_{i}{\bf{c}}_ig_i^{(0)}=0, \quad \sum_{i}{\bf{c}}_i{\bf{c}}_ig_i^{(0)}=\hat c_s^2T{\bf{I}},\quad
\sum_{i}S_i= \rho c_v\frac{\Delta t}{2} \partial_t^2T, \quad  \sum_{i}{\bf{c}}_iS_i=0,
\sum_{i}G_i=Q, \sum_{i}{\bf{c}}_iG_i=0.
\label{eq33}
\end{equation}
Summing Eq. (\ref{eq30}) and Eq. (\ref{eq32}) over $i$ with the help of Eq. (\ref{eq33}), and note that $d_{1i}^2 = {\nabla _1}{\nabla _1}:{\bf{I}}$, we obtain
\begin{equation}
\rho c_v\partial_{t_1}T= { Q^{(1)}}
 \label{eq34}
\end{equation}
\begin{equation}
 \nabla_{1}\cdot \sum_{i} c_ig_i^{(1)} +\rho c_v\partial_ {t_2}T+ \frac{\Delta t}{2}\nabla _1\nabla_1:\hat c_s^2 T{\bf{I}}=0
\label{eq35}
\end{equation}	
In order to evaluate $\sum_{i} c_ig_i^{(1)}$ in Eq. (\ref{eq35}), we multiply Eq. (\ref{eq30}) by ${{\bf{c}}_i}$ and take summation over $i$,
\begin{equation}
\begin{array}{l}
- \frac{1}{{{\tau _g}\Delta t}}\sum\limits_i {{{\bf{c}}_i}g_i^{\left( 1 \right)}}  = \rho {c_v}{\partial _{{t_1}}}\sum\limits_i {{{\bf{c}}_i}g_i^{\left( 0 \right)} + } {\nabla _1} \cdot \sum\limits_i {{{\bf{c}}_i}{{\bf{c}}_i}g_i^{\left( 0 \right)}}  - \sum\limits_i {{{\hat w}_i}{{\bf{c}}_i}{Q^{\left( 1 \right)}}}  \\
\quad \quad \quad \quad \quad \; \; 	= \hat c_s^2{\nabla _1}T.
\end{array}	
\label{eq36}
\end{equation}
Substituting Eq. (\ref{eq36}) into Eq. (\ref{eq35}) yields
\begin{equation}
\rho c_v\partial_ {t_2}T=\nabla_{1}\cdot \left(	{\tau_g}-\frac{1}{2}\right) \Delta t \hat c_s^2 \nabla_{1}T.
\label{eq37}   
\end{equation}
Based on the equations of Eqs. (\ref{eq34}) and (\ref{eq37}), we obtain 
\begin{equation}
\rho c_v\partial_t T=\nabla \cdot \left({\tau_g}- \frac{1}{2} \right)\hat c_s^2 \Delta t\nabla  T+ Q.
\label{eq38}
\end{equation}
Comparing Eq. (\ref{eq38}) with Eq. (\ref{eq30}), dimensionless relaxation time $\tau_g$ is determined by $	\lambda=\hat c_s^  2 \left({\tau_g}- \frac{1}{2} \right)\Delta t$.

Through the above multi-scale analysis, it is clear that the temperature equation is recovered without any deviation terms.  In addition, although a temperature space derivative appears in the discrete source term of $G_{i}$,  it can be computed locally using Eq. (\ref{eq36}) with $\epsilon g_i^{\left( 1 \right)} \approx {g_i} - g_i^{\left( {eq} \right)}$. More importantly, the calculations of  $\nabla  \cdot \left( {\kappa \nabla T} \right)$  or   $\nabla \left( {\rho {{\rm{c}}_v}} \right)$ do not appear in the present scheme, such that the formulation of the present LBM is more concise in contrast to previous models, and therefore, it holds the advantages of the original LBM very well. 

\subsubsection{MRT model for liquid-vapor phase change}
We now turn to extend the above BGK model to the MRT version, and the evolution equation with a MRT collision operator ${{\bf{\hat M}}}$  can be expressed as 
\begin{equation}
	\rho c_v{g_i}\left( {{\bf{x}} + {{\bf{c}}_i}\Delta t,t + \Delta t} \right) =
	{g_i}\left( {{\bf{x}},t} \right)  +(\rho c_v-1){g_i}\left( {{\bf{x}} + {{\bf{c}}_i}\Delta t,t} \right)
	-\left( {{{{\bf{\hat M}}}^{ - 1}}\Lambda {\bf{\hat M}}} \right)_{ij}\left[
	{{g_j}\left( {{\bf{x}},t} \right) - g_j^{(eq)}\left( {{\bf{x}},t}
		\right)} \right]+\Delta t {G_i} + \Delta t {S_i}  ,
\label{eq39}
\end{equation} 
where $\Lambda$ is a non-negative diagonal relaxation matrix given by $\Lambda  = \left( {{\varsigma _0},{\varsigma _1}, \ldots ,{\varsigma _4}} \right)$ with ${\varsigma _i} \in \left( {0,1} \right)$.  Different from most previous MRT models based on the orthogonal transformation matrix, the present MRT scheme is constructed using a non-orthogonal one, which is more simplicity and efficiency due to it contains more zero elements than the  orthogonal transformation matrix \cite{he2019latticeBo}. For the D2Q5 lattice considered here, the non-orthogonal transformation matrix ${{\bf{\hat M}}}$ is defined as \cite{liu2016nonorth}

\begin{equation}
{\bf{\hat M}} = \left[ {\begin{array}{*{20}{c}}
			1&1&1&1&1\\
			0&1&0&{ - 1}&0\\
			0&0&1&0&{ - 1}\\
			0&1&1&1&1\\
			0&1&{ - 1}&1&{ - 1}
	\end{array}} \right].
\end{equation}

To show that the temperature equation can be recovered correctly by the present MRT model, the aforementioned multi-scale analysis is adopted again. Applying the Taylor expansion to Eq. (\ref{eq39}), and note that Eq. (\ref{eq27}), we can obtain the following equations at different order of $\epsilon$,

\begin{equation}
{\rm O}(\epsilon ^0): \quad g_i^{(0)}=g_i^{(eq)},
\label{eq41}
\end{equation}
\begin{equation}
{\rm O}(\epsilon ^1):\quad \rho c_v\partial_ {t_1}g_i^{(0)}+d_{1i}g_i^{(0)}=-\frac{1}{\Delta t} \left({{{\bf{\hat M}}}}^{-1}{\Lambda}{{{\bf{\hat M}}}}\right)_{ij} g_j^{(1)} + {{\hat w}_i Q^{(1)}},
\label{eq42}
  \end{equation}
\begin{equation}
{\rm O}(\epsilon ^2): \quad \rho c_vD_{1i}g_i^{(1)}+\rho c_v\partial_ {t_2}g_i^{(0)}+\rho c_v \frac{\Delta t}{2}D_{1i}^2g_i^{(0)}+(1-\rho c_v)d_{1i}g_i^{(1)}+(1-\rho c_v)\frac{\Delta t}{2} d_{1i}^2g_i^{(0)}=-\frac{1}{\Delta t} \left({{{\bf{\hat M}}}}^{-1}{\Lambda}{{{\bf{\hat M}}}}\right)_{ij}g_j^{(2)}+{\hat w}_i \rho c_v\frac{\Delta t}{2} \left(\partial t_1\right)^2T.
\label{eq43} 
\end{equation}

Multiplying ${{\bf{\hat M}}}$ on both sides of of the above equations, we can obtain the following equations in the moment space, 

\begin{equation}
{\rm O}(\epsilon ^0):\quad {{{\bf{\hat m}}}}^{(0)}={{{\bf{\hat m}}}}^{(eq)},
\label{eq44} 
\end{equation}
\begin{equation}
{\rm O}(\epsilon ^1):\quad \rho c_v {\bf{I}} \partial_ {t_1}{{{\bf{\hat m}}}}^{(0)}+{\bf{d}}_{1}{{{\bf{\hat m}}}}^{(0)}=-\frac{1}{\Delta t} {\Lambda}{{{\bf{\hat m}}}} ^{(1)} + {{{\bf{\hat M}}}}{{{\bf{\hat G}}}}^{(1)},
\label{eq45} 
\end{equation}
\begin{equation}
O(\epsilon ^2): \quad \rho c_v{\bf{I}}\partial_{t_1}  {{{\bf{\hat m}}}}^{(1)}+  {\bf{d}}_{1}{{{\bf{\hat m}}}}^{(1)} +\rho c_v{\bf{I}}\partial_ {t_2} {{{\bf{\hat m}}}}^{(0)}+\rho c_v{\bf{I}} \frac{\Delta t}{2}\partial_{t_1}^2{{{\bf{\hat m}}}}^{(0)}+\rho c_v  \frac{\Delta t}{2}\left(2\partial_{t_1}{\bf{d}}_{1}\right){{{\bf{\hat m}}}}^{(0)}+ \frac{\Delta t}{2}{\bf{d}}_{1}^2{{{\bf{\hat m}}}}^{(0)}=- \frac{1}{\Delta t} {\Lambda } {{{\bf{\hat m}}}}^{(2)}+ {\bf{M}}{{{\bf{\hat S}}}}^{(2)},
\label{eq46} 
\end{equation}
where ${{\bf{d}}_1} = {\bf{\hat M}}{\rm{diag}}\left[ {{{\bf{c}}_0} \cdot \nabla , \ldots ,{{\bf{c}}_4} \cdot \nabla } \right]{{{\bf{\hat M}}}^{ - 1}}$, ${{{\bf{\hat G}}}^{\left( 1 \right)}} = {\left[ {{{\hat w}_0}{Q^{\left( 1 \right)}}, \ldots ,{{\hat w}_4}{Q^{\left( 1 \right)}}} \right]^{\rm T}}$, ${{{\bf{\hat S}}}^{\left( 2 \right)}} = {\left[ {0.5{{\hat w}_0}\rho {c_v}\partial _{{t_1}}^2T, \ldots ,0.5{{\hat w}_4}\rho {c_v}\partial _{{t_1}}^2T} \right]^{\rm T}}$. In addition,  ${\bf{\hat m}} = {\bf{\hat Mg}}$ is the moment function, and ${{{\bf{\hat m}}}^{\left( {eq} \right)}}$ is the equilibrium function in the moment space defined as
\begin{equation}
{{{\bf{\hat m}}}^{\left( {eq} \right)}} = {\bf{\hat M}}{{\bf{g}}^{\left( {eq} \right)}} = {\left[ {T,0,0,\dot wT,0} \right]^{\rm T}}.
\label{eq47} 
\end{equation} 
According to Eq. (\ref{eq45}), we can rewrite the first-order equations in $t_1$ scale, but here we just present the first, second and third ones since only these equations are useful in deducing the macroscopic equation, 
\begin{equation}	
\rho c_v \partial_ {t_1}T= {Q}^{(1)},
\label{eq48}
\end{equation}
\begin{equation}
c_s^2\partial _{x1}T =-\frac{1}{\Delta t} {\varsigma}_1{m}_1 ^{(1)},
\label{eq49}
\end{equation}
\begin{equation}
c_s^2\partial _{y1}T =-\frac{1}{\Delta t} {\varsigma}_2{m}_2 ^{(1)},
\label{eq50}
\end{equation}
Note that ${\bf{d}}_1^2{{{\bf{\hat m}}}^{\left( 0 \right)}} = \left( {{\bf{E}} \cdot {\nabla _1}} \right)\left( {{\bf{E}} \cdot {\nabla _1}} \right){{{\bf{\hat m}}}^{\left( 0 \right)}}$ where ${\bf{E}} = \left( {{{\bf{E}}_x},{{\bf{E}}_y}} \right)$, and ${{\bf{E}}_x} = {\bf{\hat M}}{\rm{diag}}\left[ {{{\bf{c}}_{0x}}, \ldots ,{{\bf{c}}_{4x}}} \right]{{{\bf{\hat M}}}^{ - 1}}$, ${{\bf{E}}_y} = {\bf{\hat M}}{\rm{diag}}\left[ {{{\bf{c}}_{0y}}, \ldots ,{{\bf{c}}_{4y}}} \right]{{{\bf{\hat M}}}^{ - 1}}$, we can also obtain the following equation for conservative variable $T$ at $t_2$ scale,
\begin{equation}
\rho {c_v}{\partial _{{t_2}}}T + {\partial _{x1}}\left\{ {{{\hat m}^{\left( 1 \right)}} + \frac{{\Delta t}}{2}\hat c_s^2{\partial _{x1}}T} \right\} + {\partial _{y1}}\left\{ {{{\hat m}^{\left( 2 \right)}} + \frac{{\Delta t}}{2}\hat c_s^2{\partial _{y1}}T} \right\} = 0.
\label{eq511}
\end{equation}
Substituting Eqs. (\ref{eq49}) and (\ref{eq50}) into Eq. (\ref{eq511}), we have  
\begin{equation}
\rho {c_v}{\partial _{{t_2}}}T = {\partial _{x1}}\left\{ {\Delta t\left( {\frac{1}{{{\varsigma _1}}} - \frac{1}{2}} \right)\hat c_s^2{\partial _{x1}}T} \right\} + {\partial _{y1}}\left\{ {\Delta t\left( {\frac{1}{{{\varsigma _2}}} - \frac{1}{2}} \right)\hat c_s^2{\partial _{y1}}T} \right\},
\label{eq53}
\end{equation} 
Based on the equations of Eq. (\ref{eq48}) and Eq. (\ref{eq53}), we can get  
\begin{equation}
\rho {c_v}{\partial _{{t}}}T = \nabla  \cdot \left( {\lambda \nabla T} \right) + Q,	
\end{equation} 
where $\lambda  = \Delta t\left( {{\varsigma _1}^{ - 1} - 0.5} \right)\hat c_s^2 = \Delta t\left( {{\varsigma _2}^{ - 1} - 0.5} \right)\hat c_s^2$.

The above procedure shows that the temperature equation can be reovered correctly from the present MRT model, and since the convection term is actually treated as a source term, the nondiagonal relaxation time used in Zhang et al.'s work is not needed. In addition, the temperature gradient appeared in the source term $Q$ can also be calculated locally in the present MRT model, and the corresponding algorithm is given by Eqs. (\ref{eq49}) and (\ref{eq50}).     

\subsection{Wetting boundary condition}
To model fluid-solid interactions, the geometric formulation proposed by Ding and Spelt \cite{ding2007wetting} is adopted in our following simulations, which can be expressed as  
\begin{equation}
{\rho _{x,0}} = {\rho _{x,2}} + \tan \left( {\frac{\pi }{2} - {\theta ^{eq}}} \right)\left| {{\rho _{x + 1,1}} - {\rho _{x - 1,1}}} \right|,
\end{equation}
where ${{\theta ^{eq}}}$ is an analytical prescribed equilibrium contact angle,  ${\rho _{x,0}}$ denotes the fluid density at the ghost layer next to the solid boundary. Here, the first index in $\rho_{x,y}$ represents the coordinate along the horizontal solid wall, while the second one denotes the coordinate normal to the solid wall. In addition, the phase interface position is defined as $\rho  = 0.5\left( {{\rho _l} + {\rho _g}} \right)$ throughout this work, in which ${\rho _l}$, ${\rho _g}$ are the liquid density and gas density, respectively.

\section{Results and discussions}
In this section, several benchmark liquid-vapor phase change problems are selected to validate the performance of our proposed LB model. These typical examples include $D^2$ law for droplet evaporation, droplet evaporation on heated surface and bubble nucleation and departure in nucleate boiling. In addition, unless otherwise specified, the aforementioned MRT model is adopted in our simulations for its good numerical accuracy and stability.  Moreover, it should be noted that apart from the relaxation times related to physical parameters, the other relaxation factors in velocity and temperature fields are set as 1.0, respectively.         

\subsection{Validation of the $D^2$ law}

\begin{figure}[tbph]
	\centering
	\subfigure{ \label{fig5a}
		\includegraphics[width=0.25\textwidth]{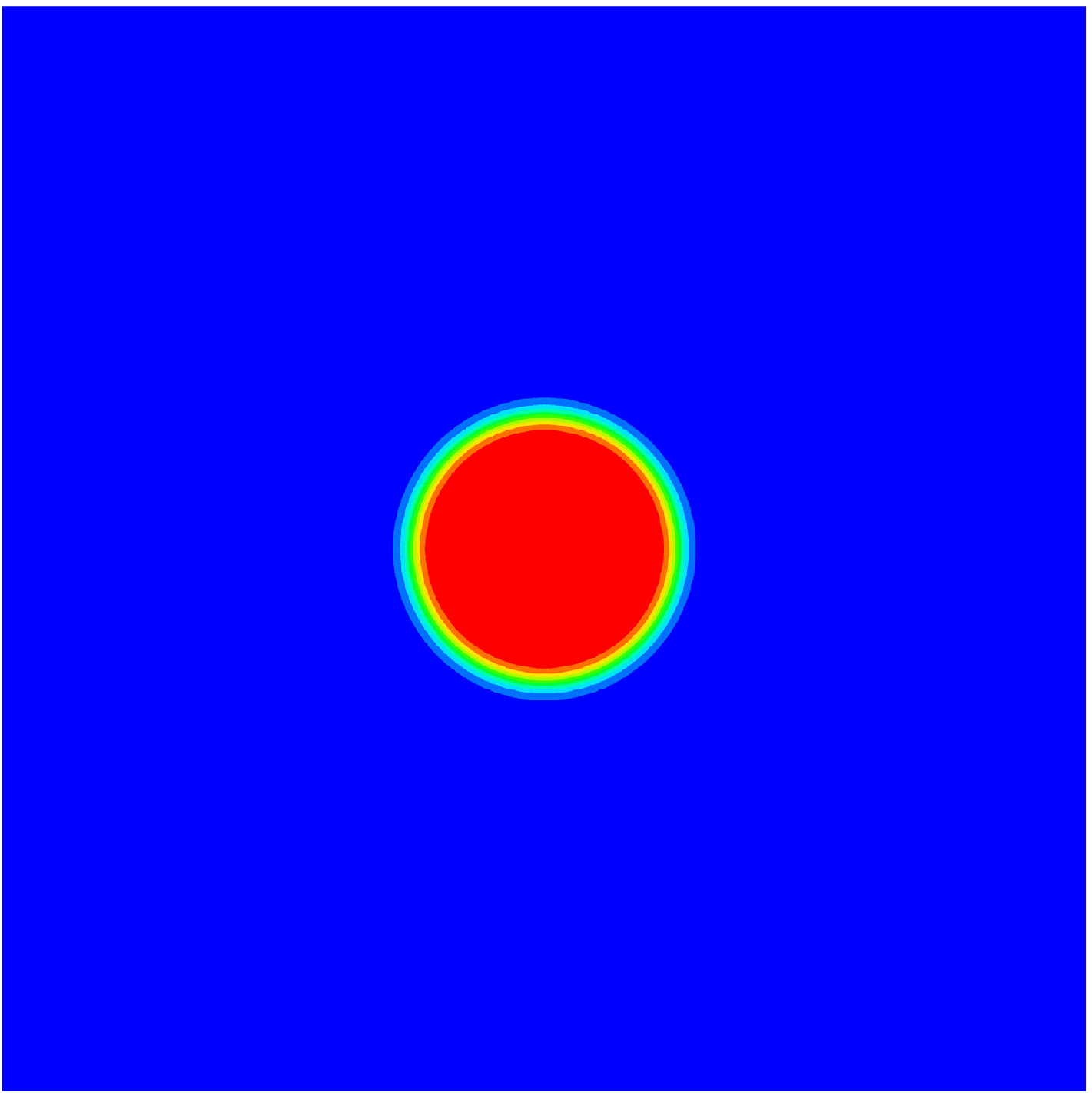}}
	\subfigure{ \label{fig5b}
		\includegraphics[width=0.25\textwidth]{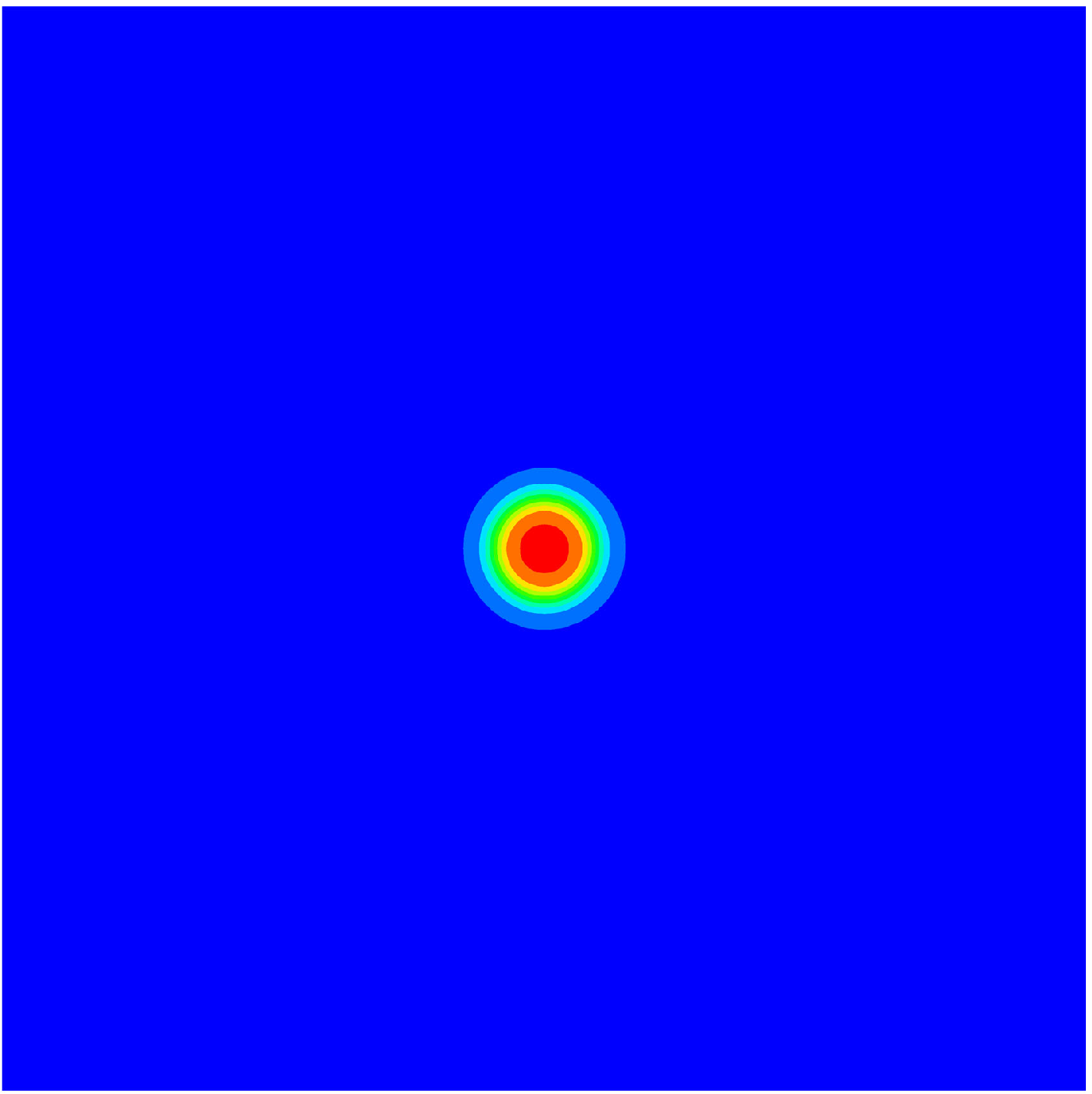}}
	\subfigure{ \label{fig5b}
	\includegraphics[width=0.25\textwidth]{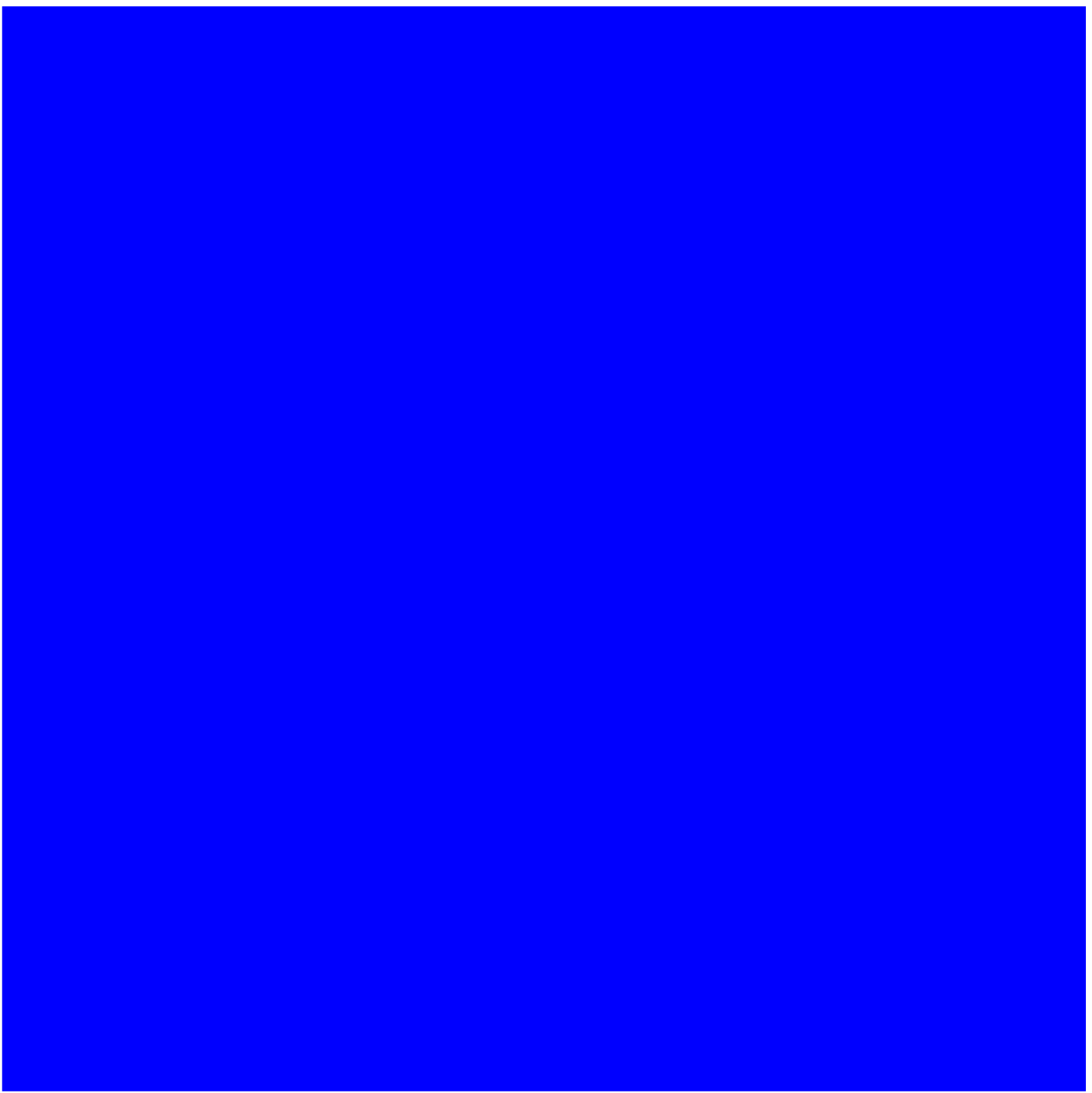}}
	\caption*{(a)}
		
	\subfigure{ \label{fig5a}
	\includegraphics[width=0.25\textwidth]{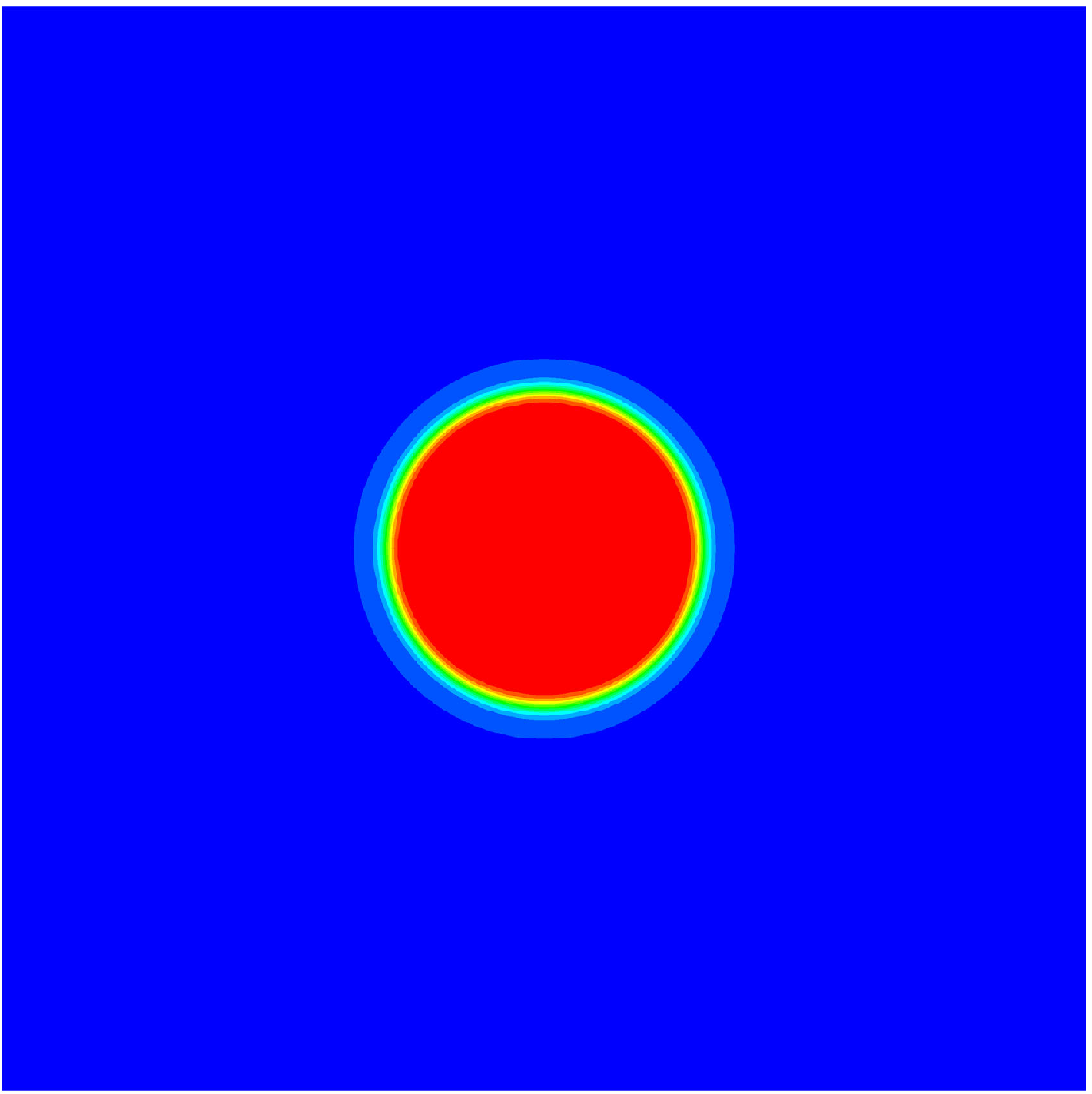}}
\subfigure{ \label{fig5b}
	\includegraphics[width=0.25\textwidth]{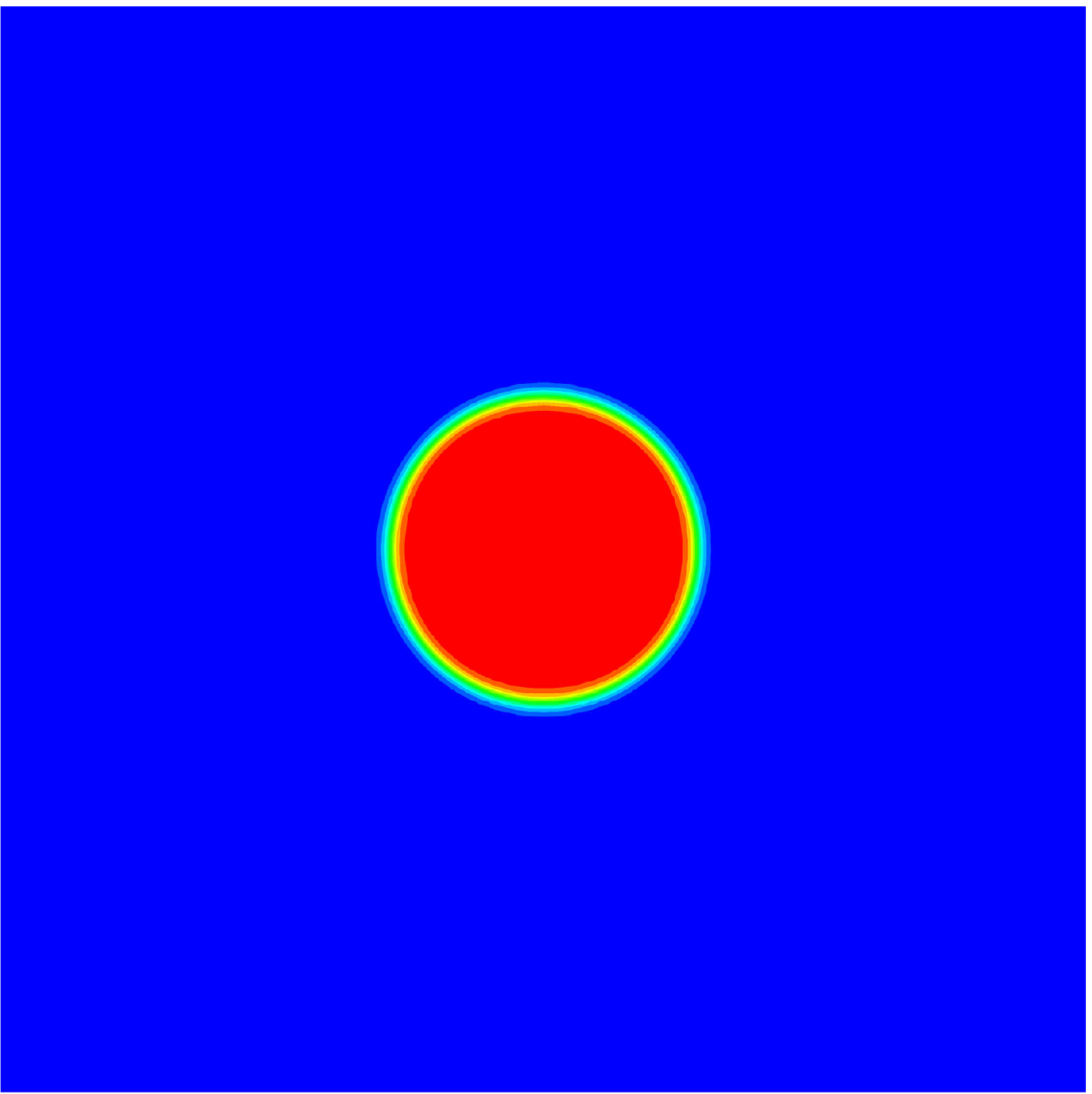}}
\subfigure{ \label{fig5b}
	\includegraphics[width=0.25\textwidth]{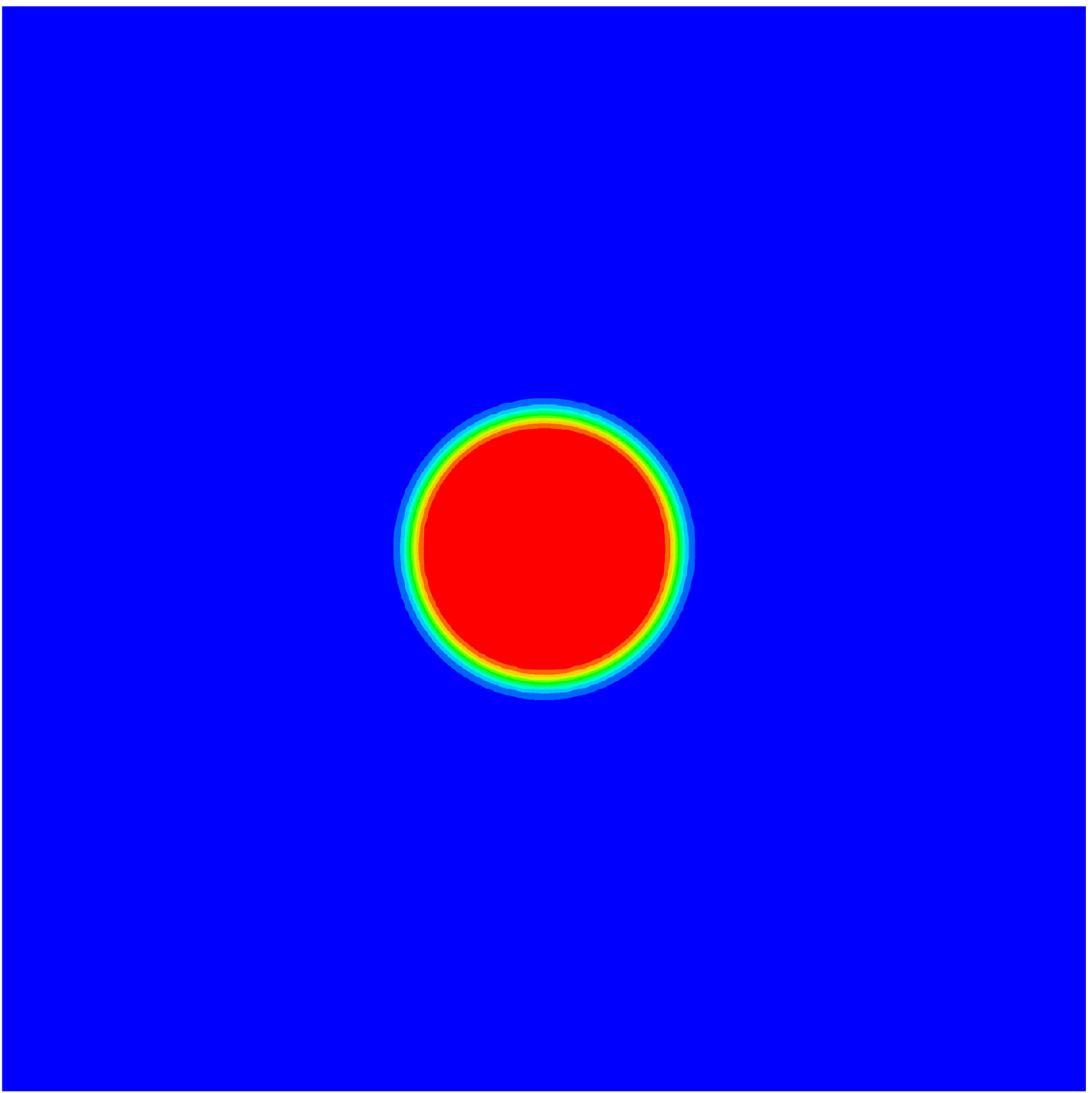}}
	\caption*{(b)}
		
	\subfigure{ \label{fig5a}
	\includegraphics[width=0.25\textwidth]{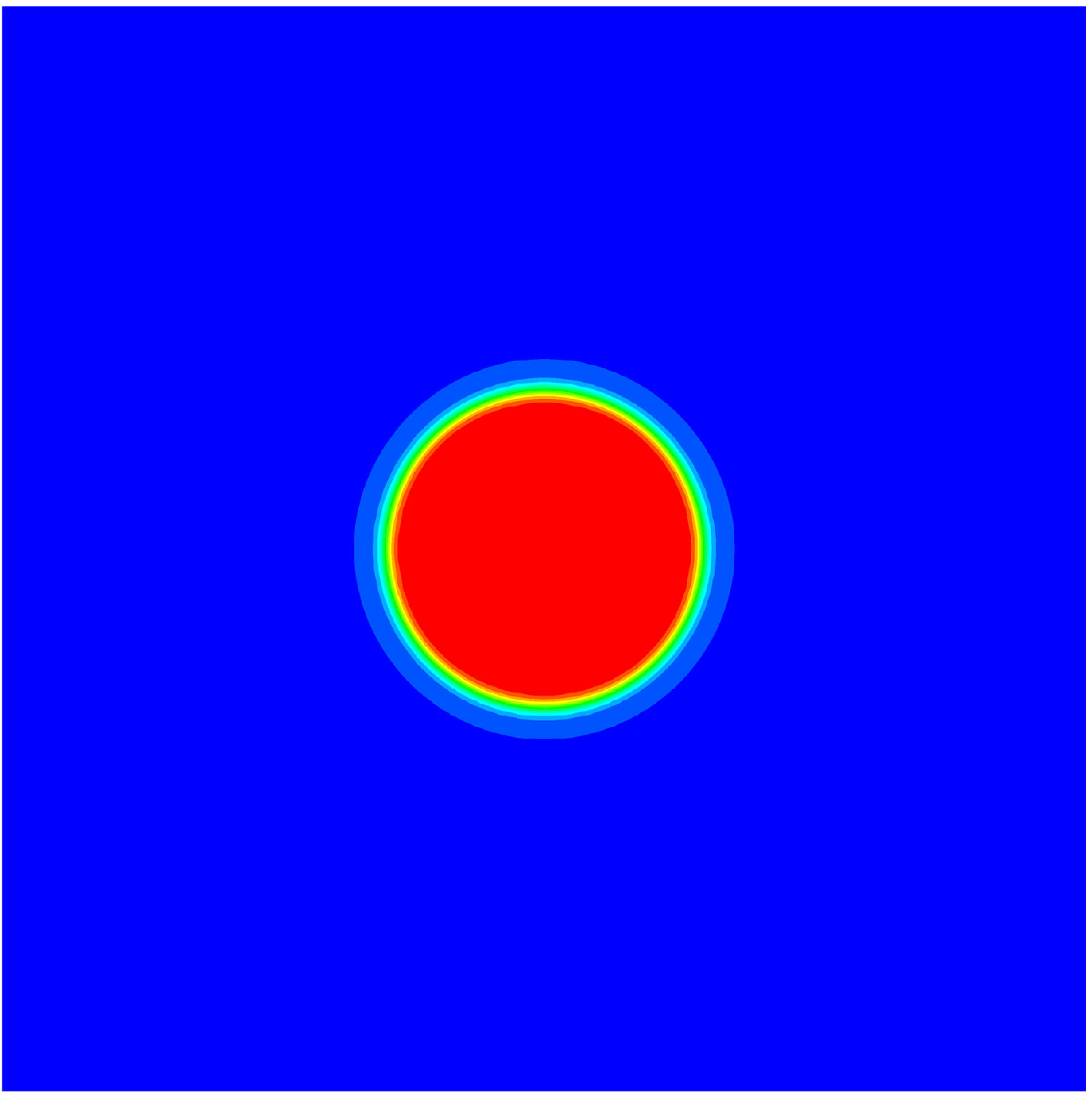}}
\subfigure{ \label{fig5b}
	\includegraphics[width=0.25\textwidth]{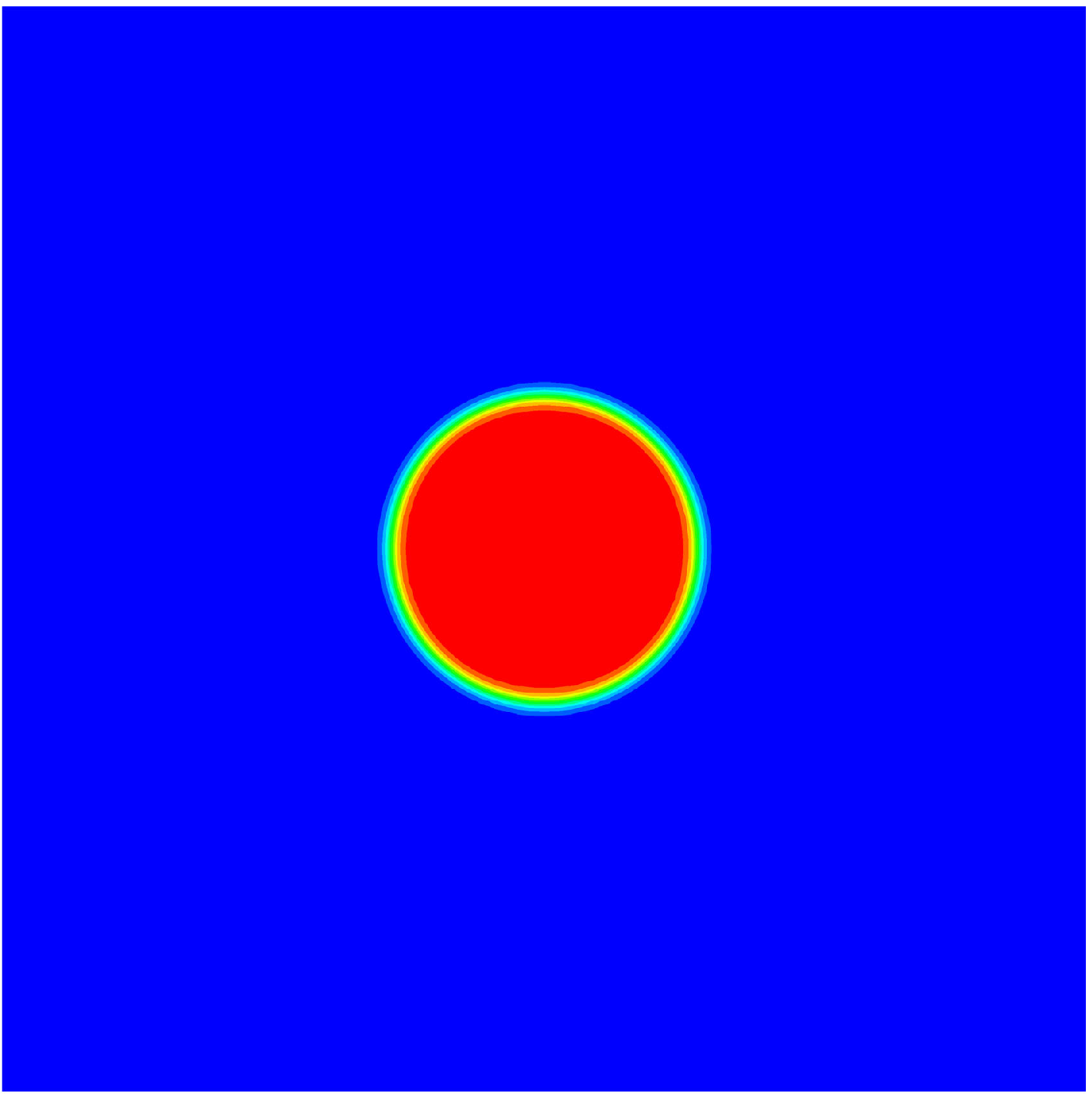}}
\subfigure{ \label{fig5b}
	\includegraphics[width=0.25\textwidth]{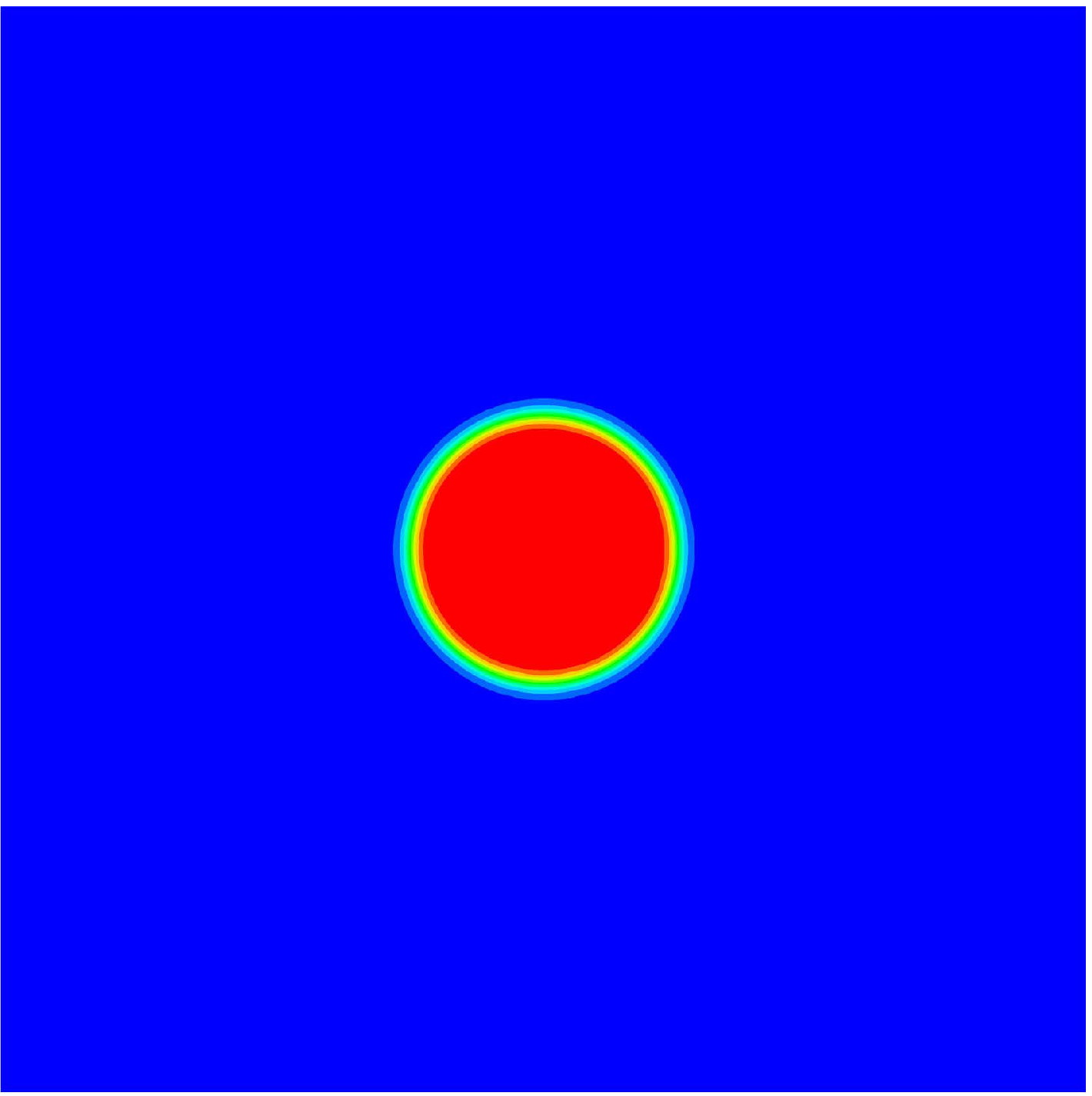}}
	\caption*{(c)}		
	\caption{Comparisons of the density contours given by the Gong-Cheng's model (a), the proposed model (b), and the WENO scheme (c) at $t^*=0.1$, $t^*=0.25$ and $t^*=0.75$ (from left to right).    }
	\label{fig1}
\end{figure}

We first validate the present thermal LB model by considering the droplet evaporation in open space, and it is well known that the variation of the droplet diameter in this problem is described by the $D^{2}$ law, which states that the square of the evaporating droplet diameter $D$ decreases linearly with time \cite{zheng2018numerical},     
\begin{equation}
	{\left( {\frac{D}{{{D_0}}}} \right)^2} = 1 - \kappa t,
\end{equation}
where $D_0$ is the droplet initial diameter and $\kappa$ is the evaporation constant. The computational domain in our simulations is a square cavity and the lattice size of it is set to be  $200 \times 200$. Initially, a droplet with a diameter of $D_0=60$ is located in the center of the cavity, and the temperature of the droplet is equal to the saturation temperature $T_{sat} = 0.86T_{c}$, while a higher temperature of $T_g=1.0Tc$ is assigned to the surrounding vapor phase. In such a case, the droplet is expected to evaporation as a result of the temperature gradient around the liquid-vapor interface.  In addition,  a Dirichlet boundary (i.e., $T=T_{g}$) is adopted for the temperature field, which can be easily established by employing the halfway bounce-back scheme. For the velocity field,  a periodic scheme is used to determain the unknown density distribution functions streaming from the outside of the boundary. The thermal conductivity, the kinematic viscosity as well as the specific heat at constant volume are chosen as $1/3$, $5.0$ and $0.1$ in the whole domain, which are all the same as previous works \cite{li2017improved,zhang2021improved}.

\begin{figure}[H]
	\centering
	\includegraphics[width=0.5\textwidth]{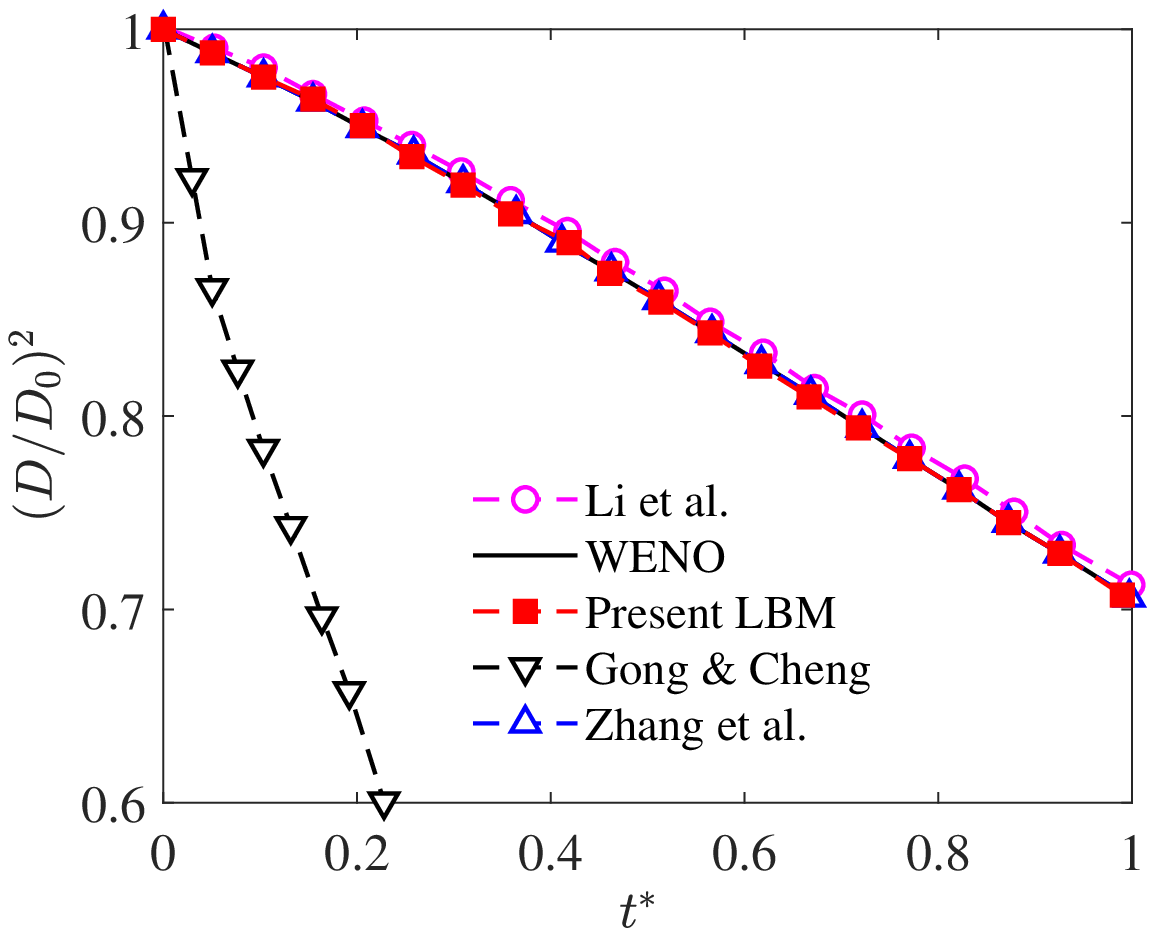}
	\caption{Time evolution of the square of the dimensionless diameter for different models.} 
	\label{fig2}
\end{figure}

Fig. \ref{fig1} presents the density contours at different dimensionless time $t^*=t/t_{total}$, and in order to test the numerical performance of the present LB model,  the results obtained from Gong-Cheng's \cite{gong2012alattice} model and the fifth-order finite difference weighted essentially non-oscillatory (WENO) scheme \cite{shu1996wneo} are also included in this figure. It is clear that the simulation results using the present LBM agree well with the WENO scheme, which suggests that the present model could provide acceptable numerical results in simulating liquid-vapor phase change.  However, it is noted that the evaporation rate predicted by the Gong-Cheng's model \cite{gong2012alattice} is much larger than the present LB and the WENO schemes, which is caused by the incorrect treatment of the temperature equation (see Set. \ref{set31} for details). To have a better understand on this statement, the time evolution of the square of the dimensionless diameter is also presented in Fig. \ref{fig2}, in which the results obtained with Li et al. 's model \cite{li2017improved} and Zhang et al.'s model \cite{zhang2021improved} are also incorporated. As seen from this figure, apart from the Gong-Cheng's model \cite{gong2012alattice}, all of the other models follow the $D^2$ law. On the other hand, although the numerical results obtained from the present model and Zhang et. al.'s model \cite{zhang2021improved} agree well with the WENO scheme, the present model is more concise as mentioned previously.   

\subsection{Droplet evaporation on heated surface}
Since the solid surface is usually encountered in most liquid-vapor phase transition problems, in the following we intend to study the problem of the droplet evaporation on heated surface, which is a standard simple test for liquid-vapor phase change. Apart from the thermal diffusivity is taken as $\eta  = 0.08$, the set-up of the other  physical parameters are the same as the first example. The  physcial domain is a rectangular enclosure which is covered by $100\times200$ mesh points. The bottom wall is a solid wall with a contact angle of ${90^ \circ }$, and it is maintained at temperature of $T_h$ in the simulations, while  the temperature for the top wall is taken as $T_s$ . The open boundary condition and the periodic boundary condition are imposed at the top wall and the horizontal direction, respectively. Initially, the density and velocity are set according to  
\begin{equation}
\rho  = \frac{{{\rho _l} + {\rho _g}}}{2} - \frac{{{\rho _l} - {\rho _g}}}{2}\tanh \left( {\frac{{\left| {{\bf{x}} - {{\bf{x}}_0}} \right| - {R_0}}}{{0.5{w_{\lg }}}}} \right),
\end{equation}
\begin{equation}
{\bf{u}}\left( {{\bf{x}},0} \right)=\bf{0},
\end{equation}
where ${{\bf{x}}_0} = {\left( {100\Delta x,0} \right)^{\rm T}}$, ${R_0} = 35\Delta x$ and ${{w_{\lg }}}$ is chosen as $5\Delta x$. The simulation is first conducted without evaporation until the contact angle of the droplet equals to the prescribed value.  
\begin{figure}[H]
	\centering
	\subfigure{ \label{fig5a}
		\includegraphics[width=0.3\textwidth]{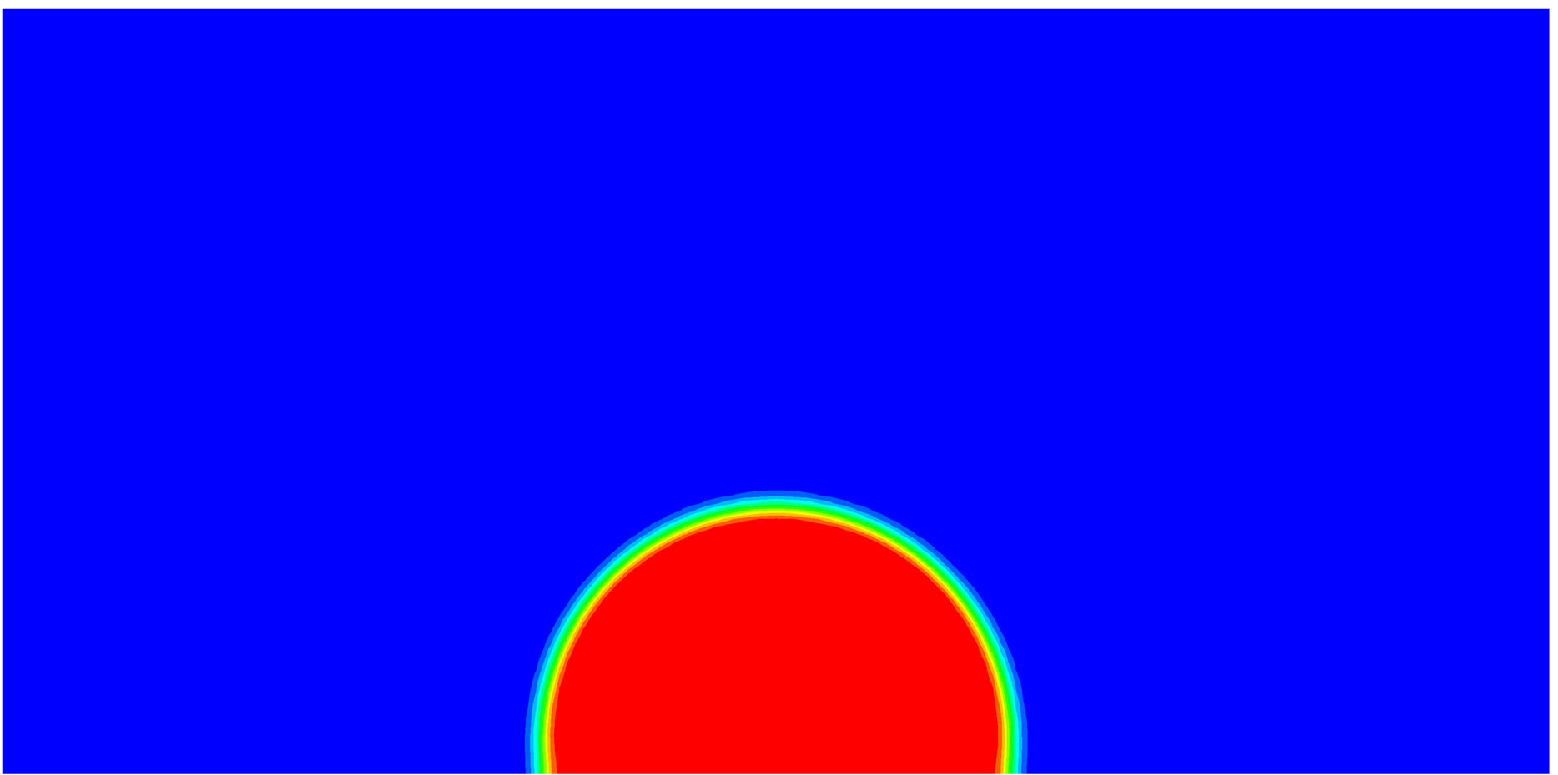}}
	\subfigure{ \label{fig5b}
		\includegraphics[width=0.3\textwidth]{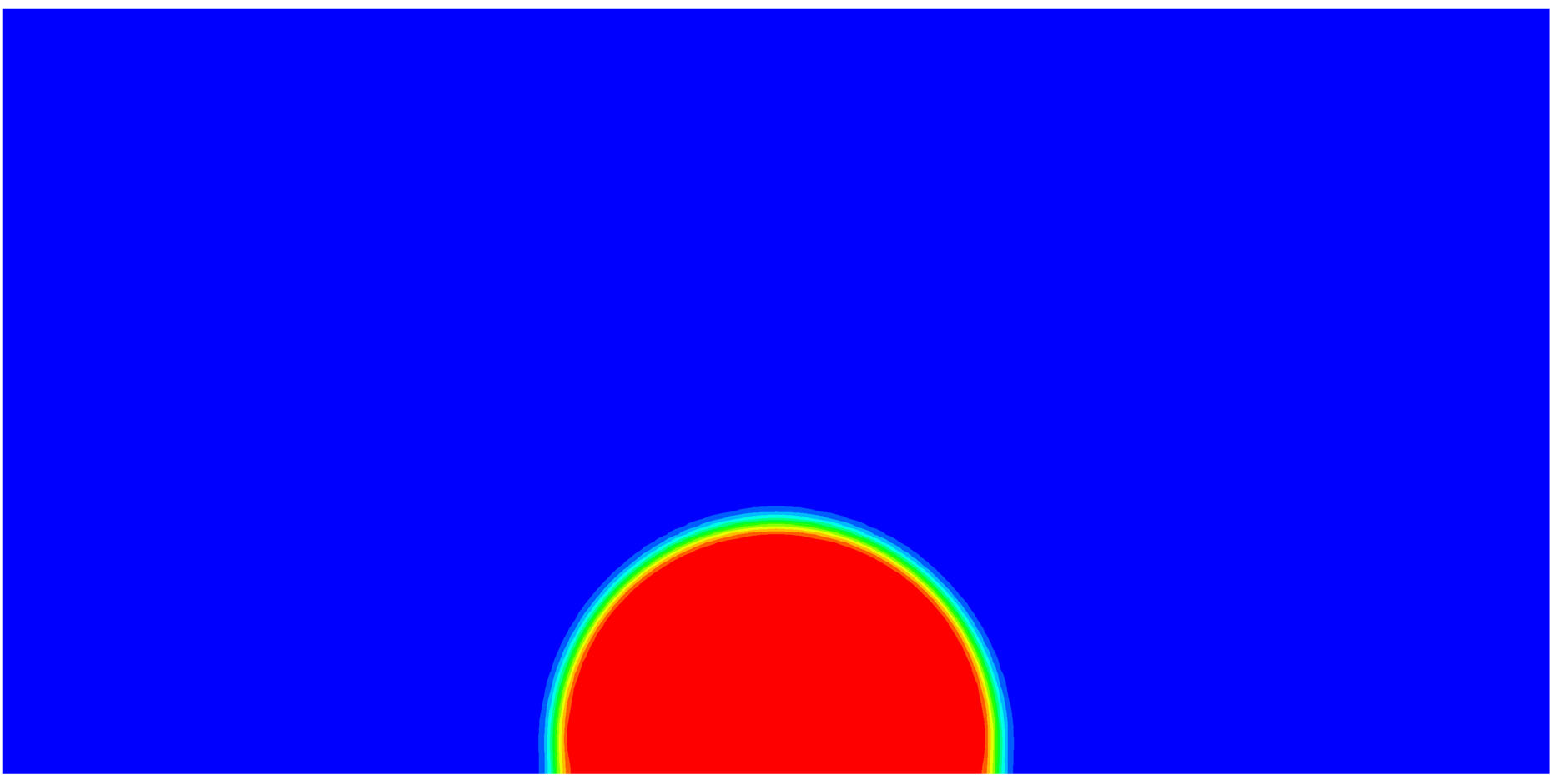}}
	\subfigure{ \label{fig5b}
		\includegraphics[width=0.3\textwidth]{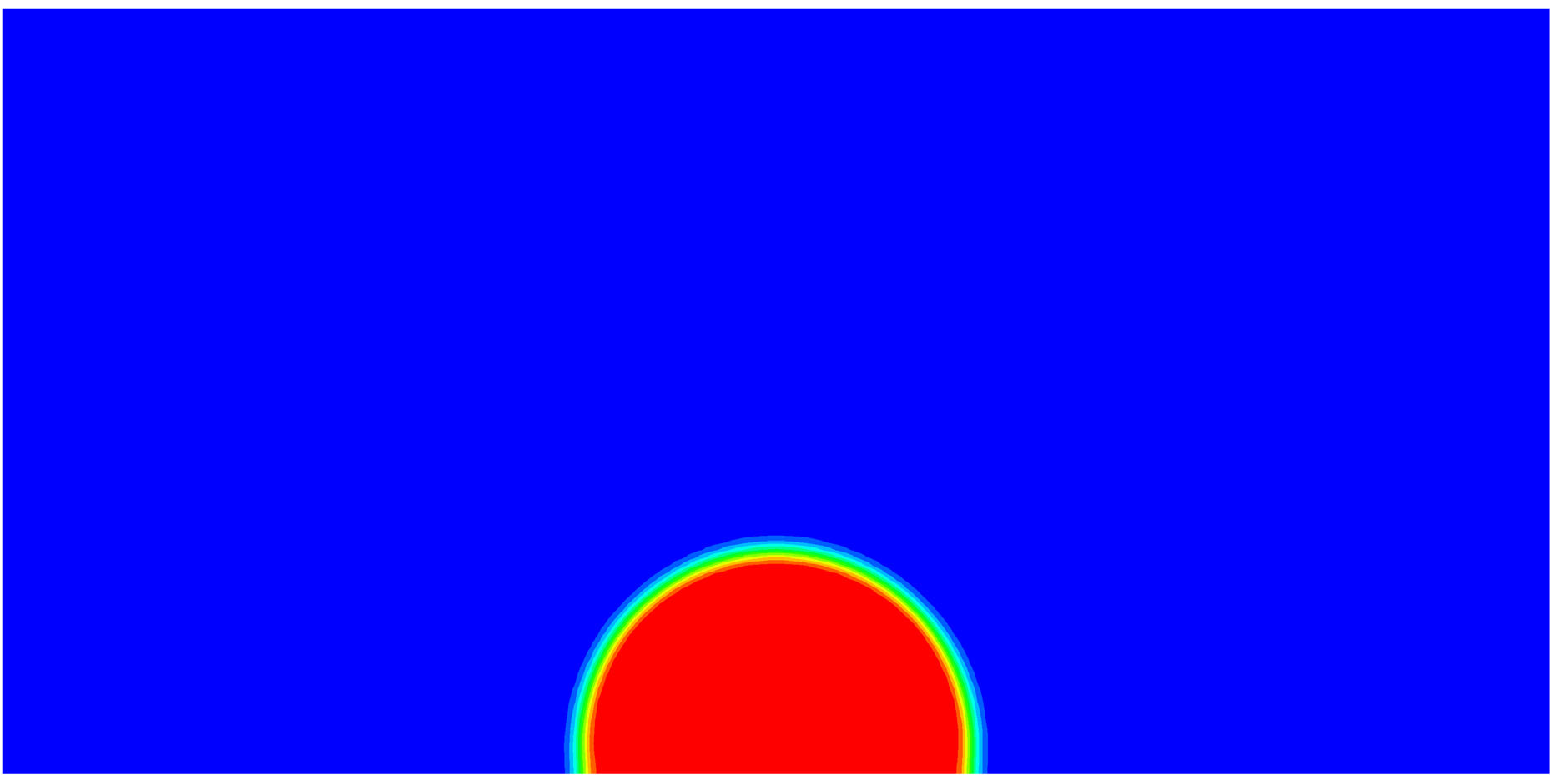}}
		\caption*{(a)}
			
	\subfigure{ \label{fig5a}
		\includegraphics[width=0.3\textwidth]{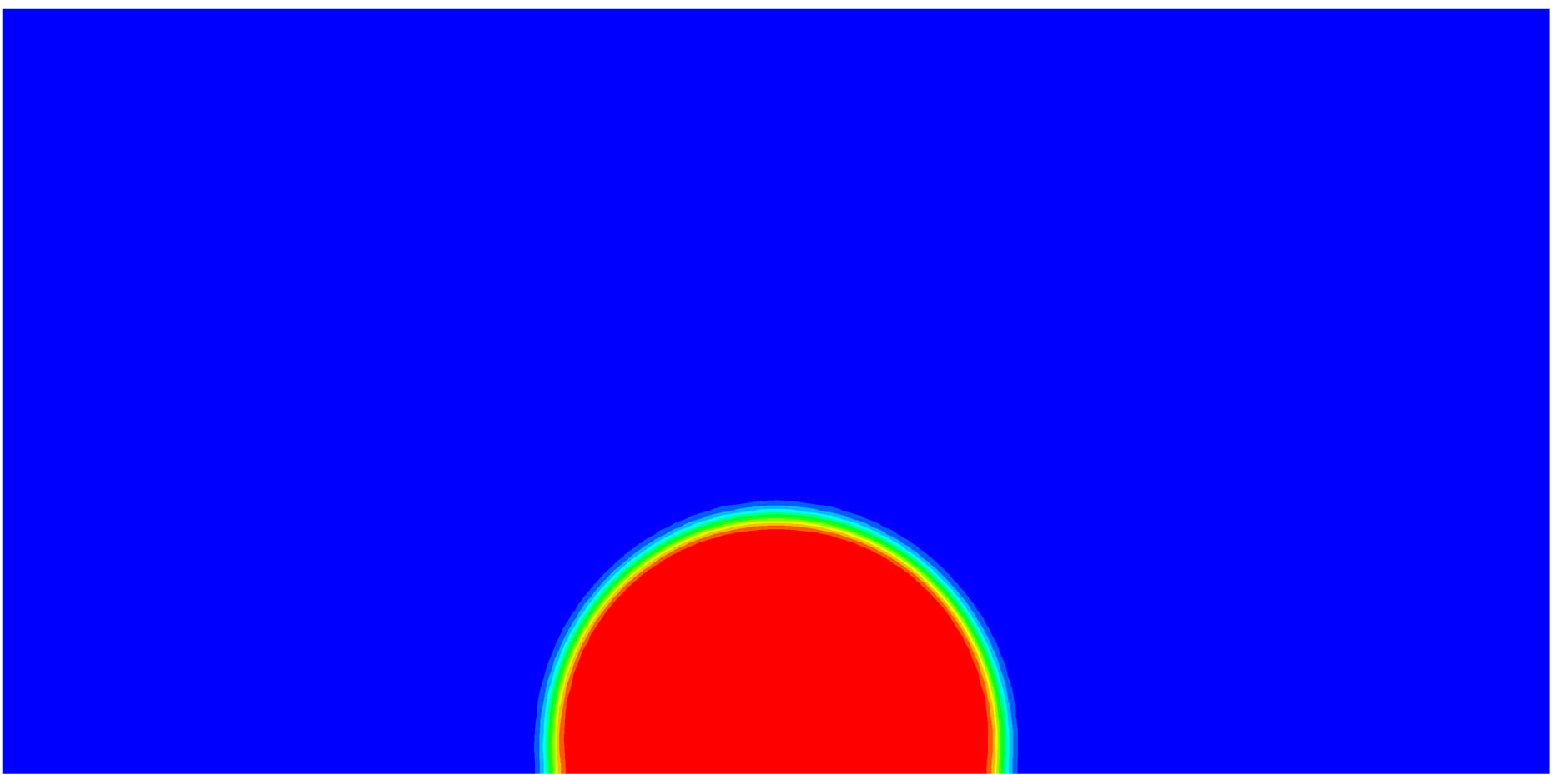}}
	\subfigure{ \label{fig5b}
		\includegraphics[width=0.3\textwidth]{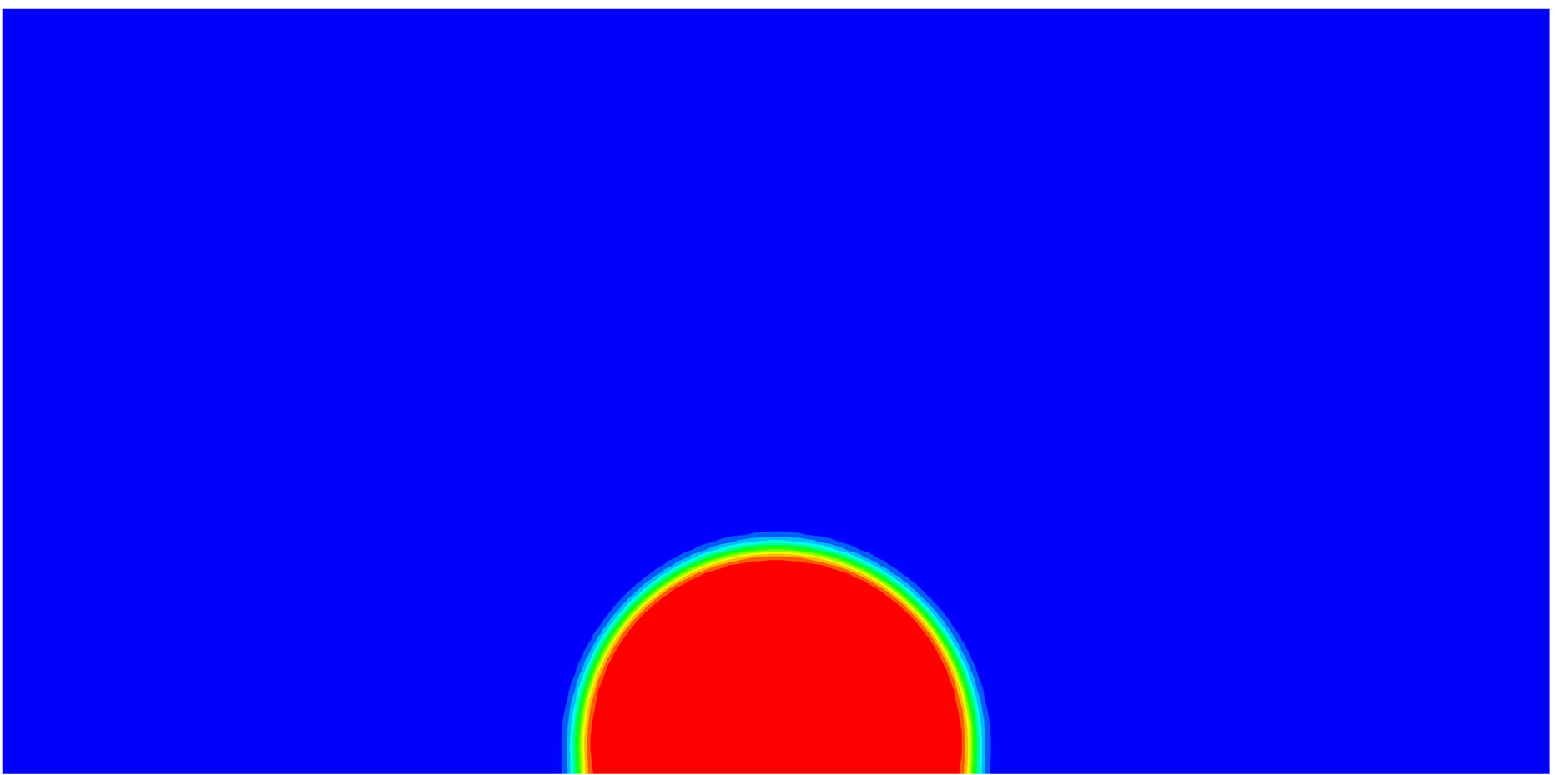}}
	\subfigure{ \label{fig5b}
		\includegraphics[width=0.3\textwidth]{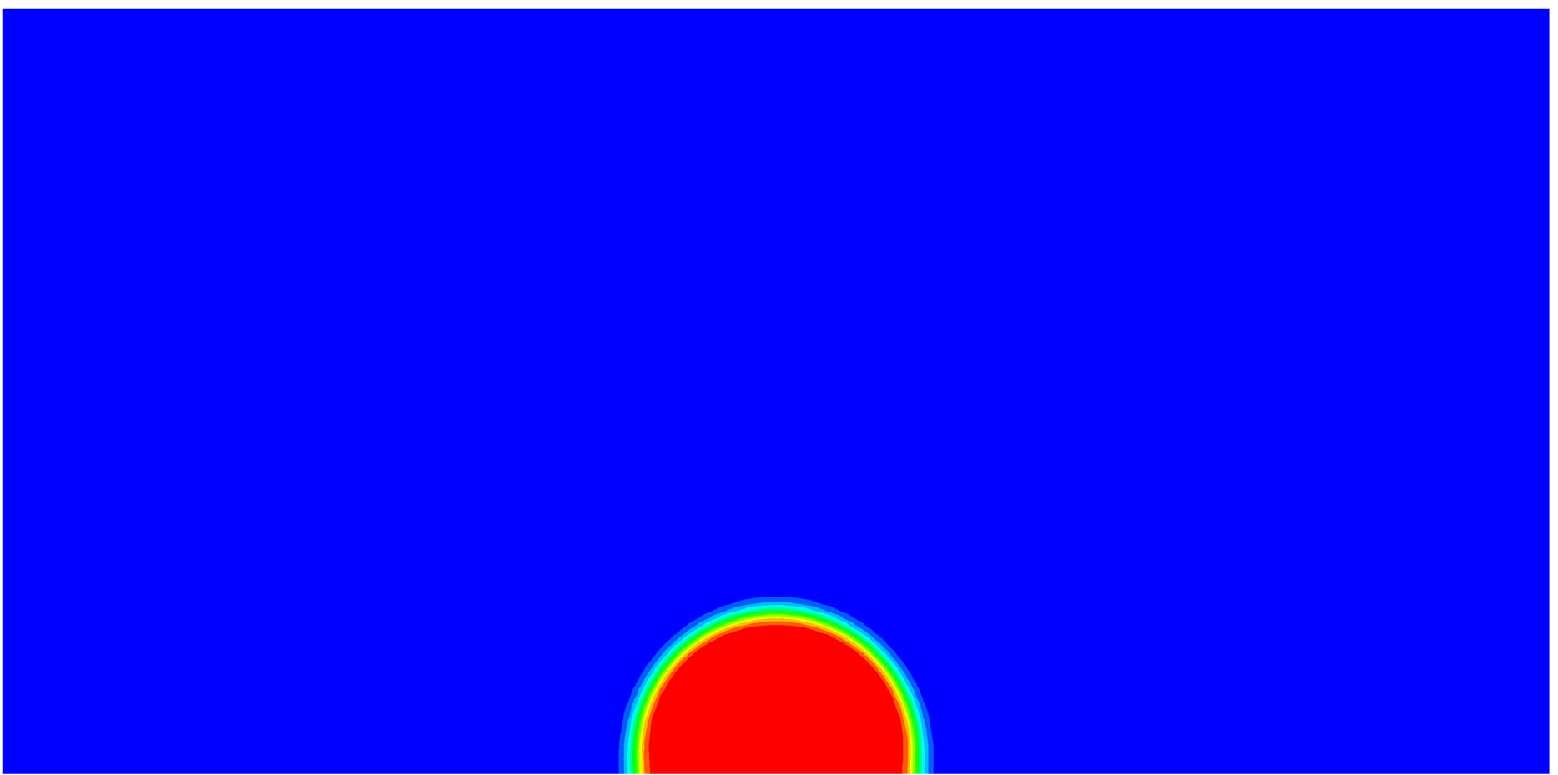}}
		\caption*{(b)}
			
	\subfigure{ \label{fig5a}
		\includegraphics[width=0.3\textwidth]{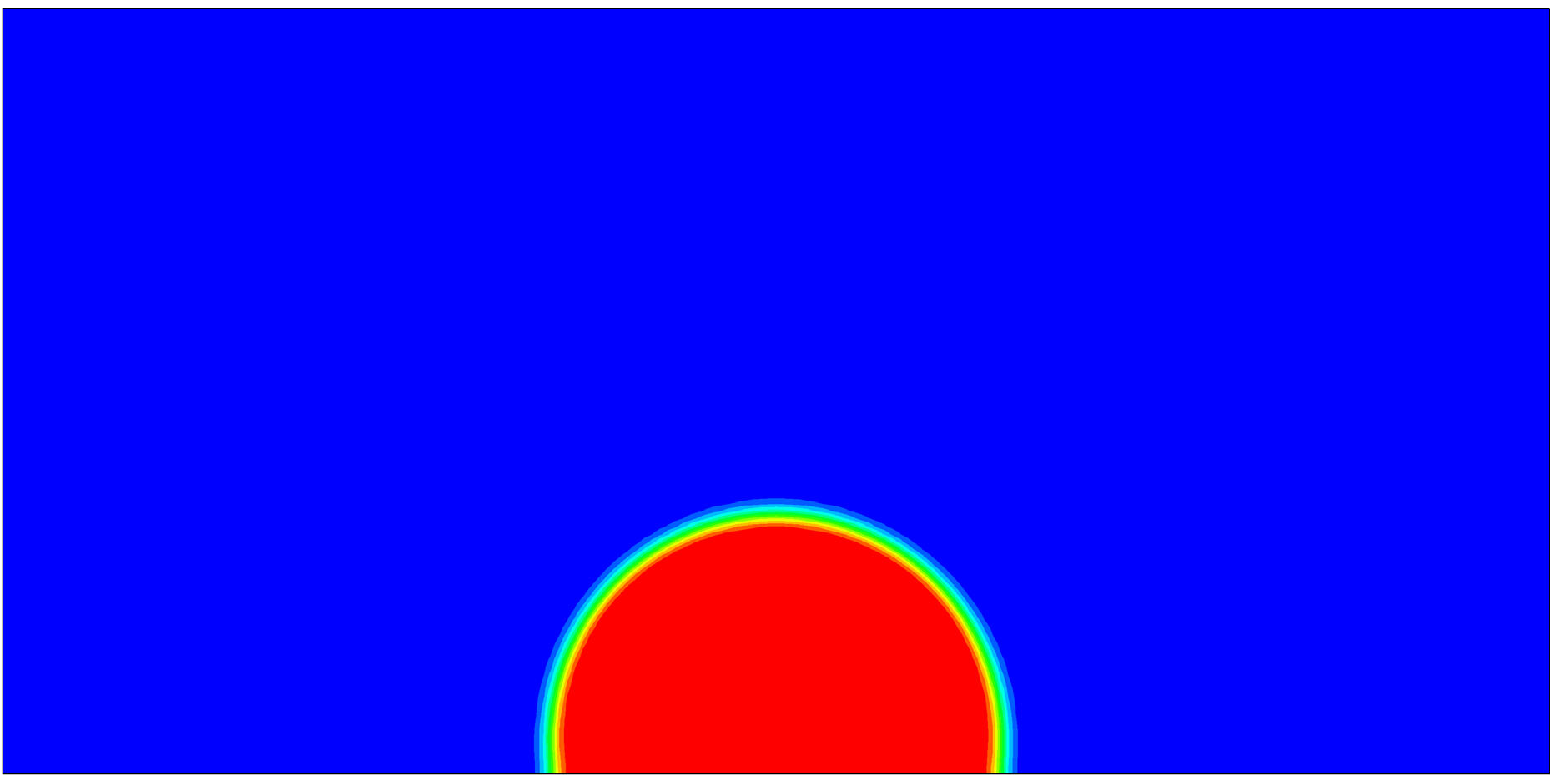}}
	\subfigure{ \label{fig5b}
		\includegraphics[width=0.3\textwidth]{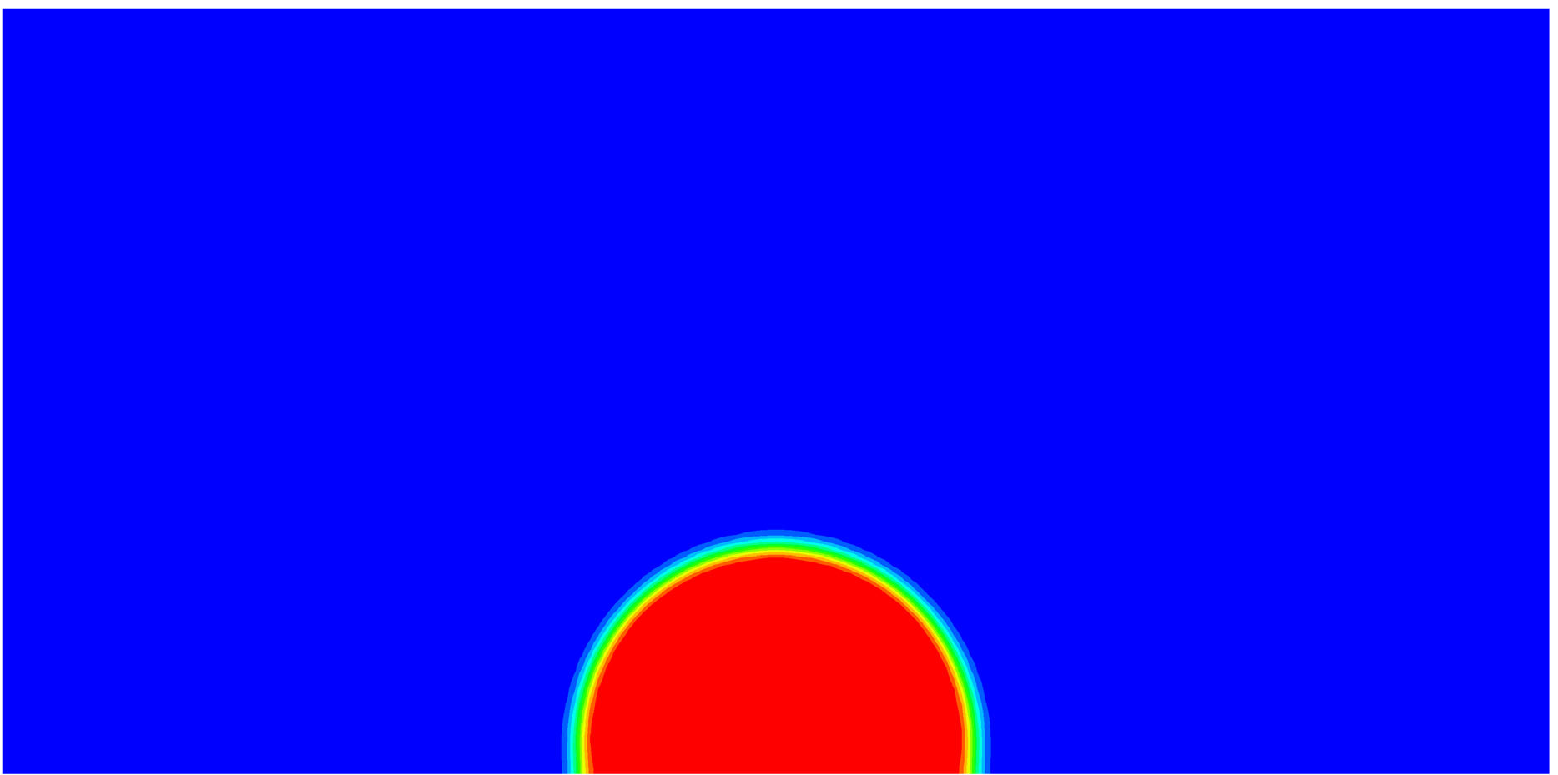}}
	\subfigure{ \label{fig5b}
		\includegraphics[width=0.3\textwidth]{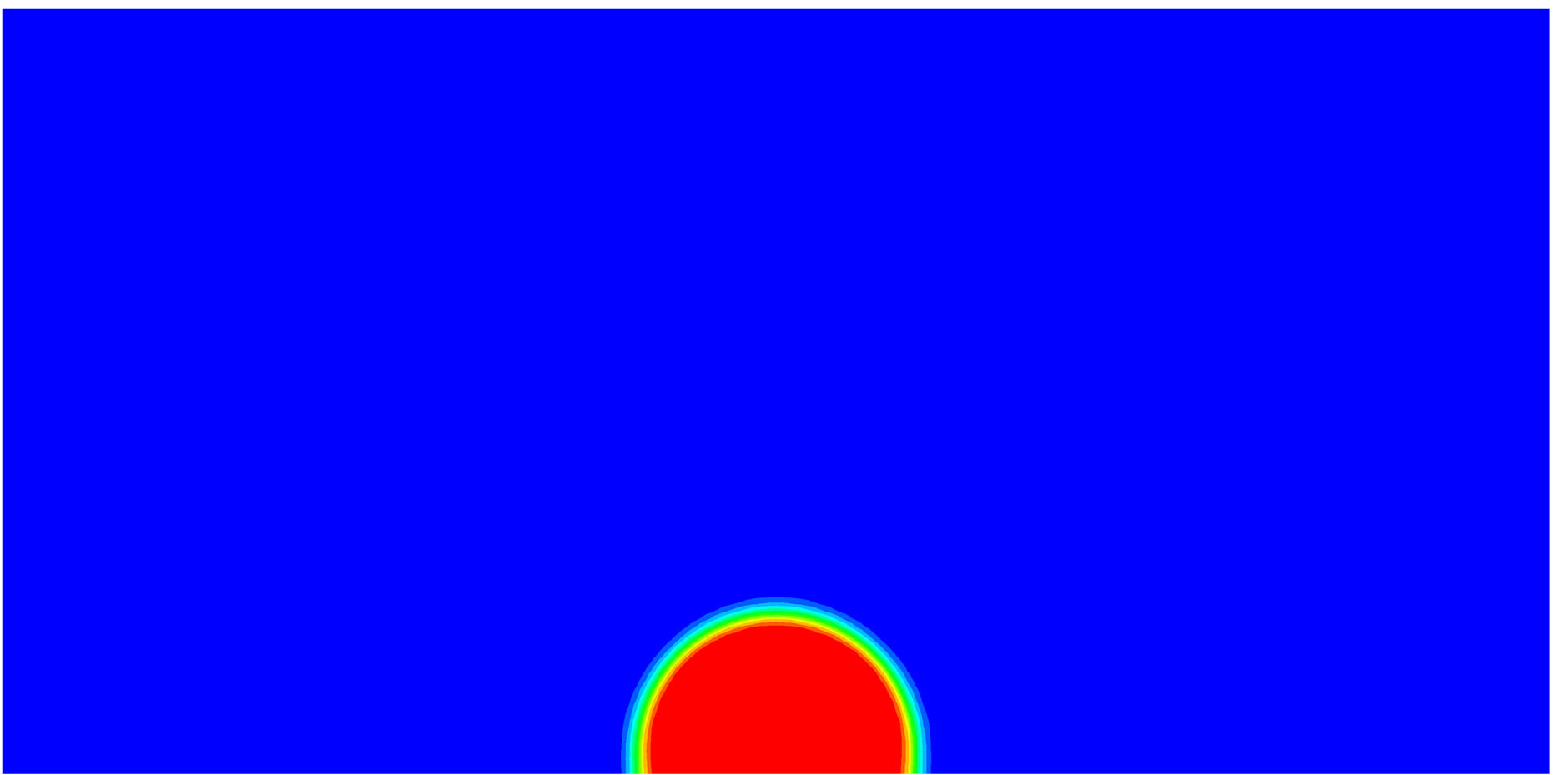}}
		\caption*{(c)}
				
	\caption{Comparisons of the density contours given by the Gong-Cheng's model (a), the proposed model (b), and the WENO scheme (c) at $t^*=0.1$, $t^*=0.25$ and $t^*=0.75$ (from left to right).    }
	\label{fig3}
\end{figure}

\begin{figure}[H]
	\centering
	\includegraphics[width=0.5\textwidth]{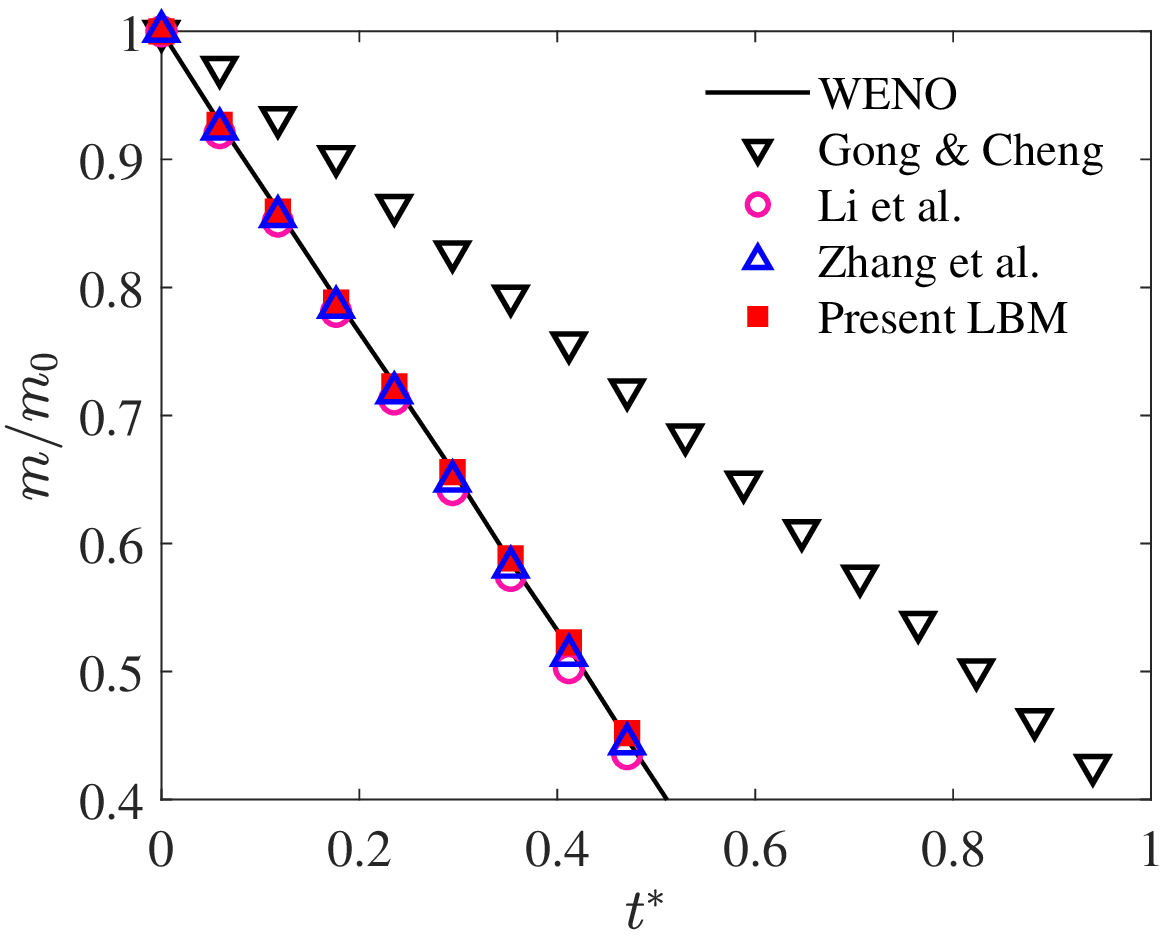}
	\caption{Time evolution of the dimensionless droplet mass for different models.} 
	\label{fig4}
\end{figure}

The comparisons of density contours among Gong-Cheng, present  and WENO solutions are shown in Fig. \ref{fig3}, where the density distributions from our new and WENO schemes match very very well, but the
evaporation process predicted by Gong-Cheng's model \cite{gong2012alattice} is a little bit slower than the other two models, and we note that the similar phenomenon is also reported by Li et al. \cite{li2017improved} and Zhang et al. \cite{zhang2021improved}. In order to further validate our new model,  the variations of the normalized droplet mass for different models are also depicted in Fig. {\ref{fig4}}. It is seen that apart from the Gong-Cheng's model \cite{gong2012alattice}, the other solutions have excellent agreement. 
 
\subsection{Bubble nucleation and departure}  
\begin{figure}[H]
	\centering
	\subfigure[]{ \label{fig5a}
		\includegraphics[width=0.25\textwidth]{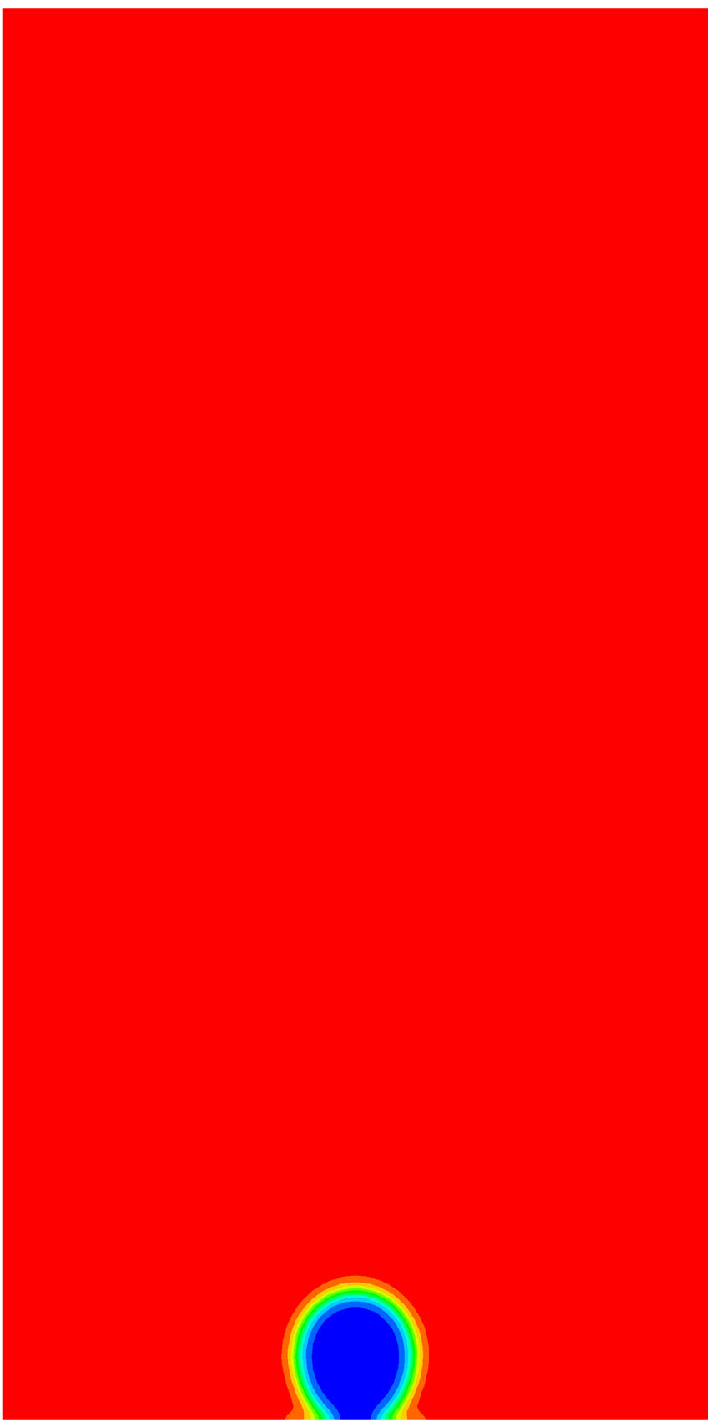}}
	\subfigure[]{ \label{fig5b}
		\includegraphics[width=0.25\textwidth]{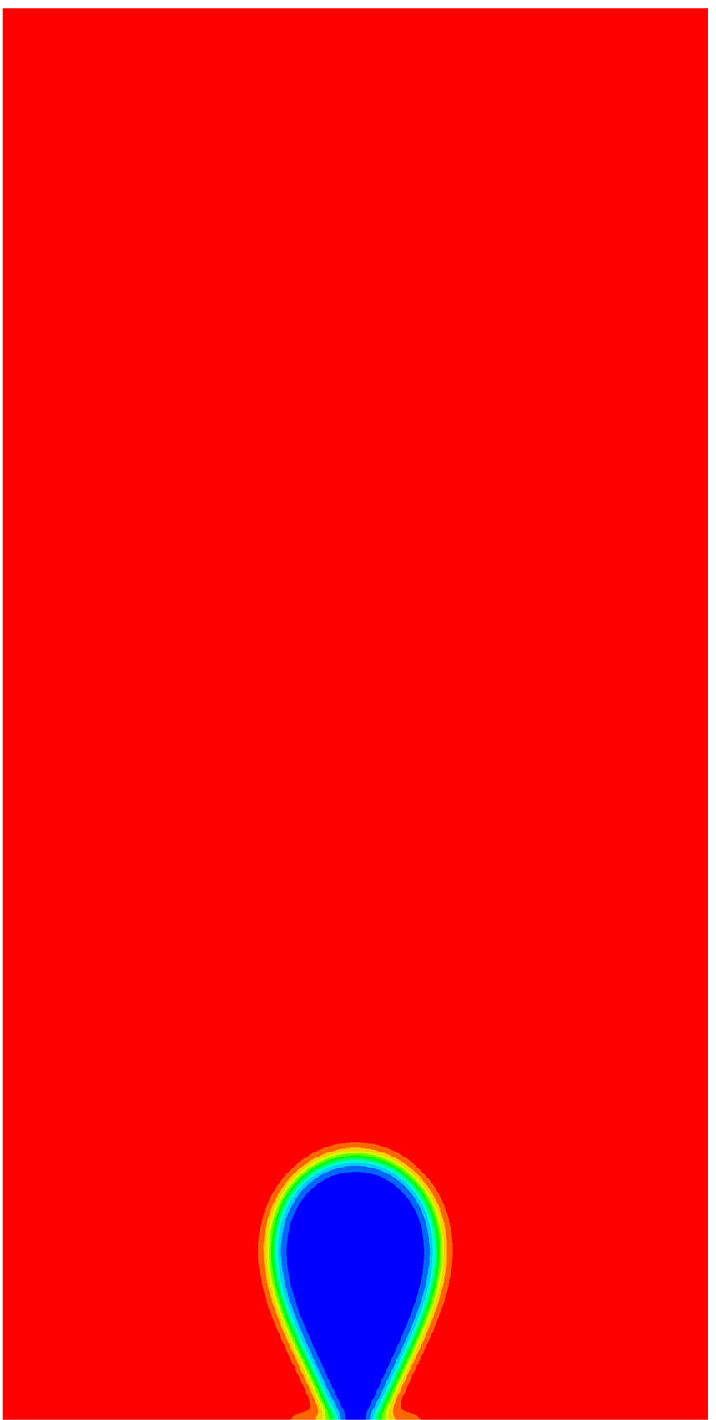}}
	\subfigure[]{ \label{fig5b}
		\includegraphics[width=0.25\textwidth]{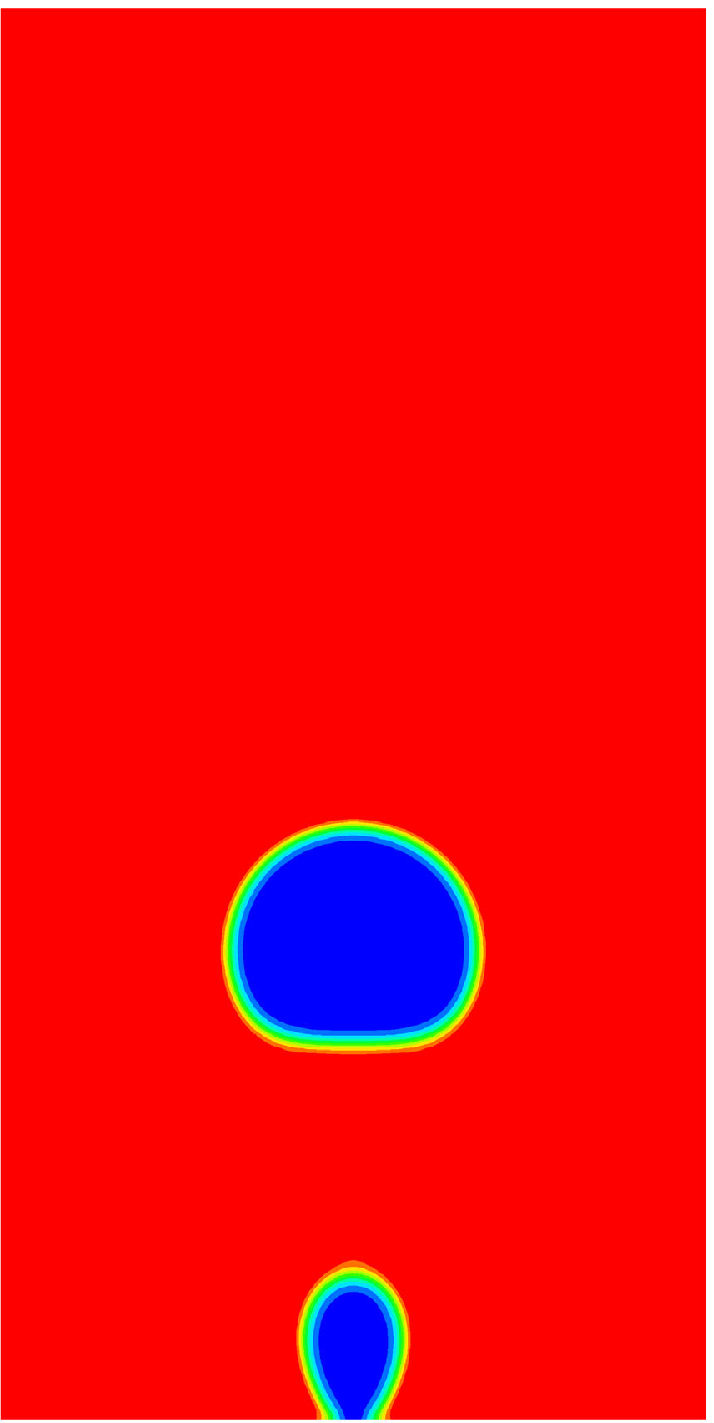}}\\		
	\caption{The distributions of the  density contours in nucleate boiling at $t=5000 \Delta t$ (a), $t=10000 \Delta t$  (b) and $t=16000 \Delta t$ (c) [$g=3.0 \times 10^{-5}$].    }
	\label{fig5}
\end{figure}

At last, to show the potentional of the present model, we consider a extremely complex liquid-vapor phase change problem of bubble nucleation and departure in nucleate boiling.  In our simulation, the computational mesh is a $150 \times 300$ rectangular domain with the periodic boundary condition in the horizontal direction.  Initially, the enclosure is filled with the saturated water, and the temperature within the domain is set to $T_{sat}=0.86T_{c}$. For the velocity field, the bottom surface is the solid wall imposed by the no-slip boundary condition, while the open boundary condition \cite{lou2013pre} is used for the top plane. Apart from the central five grids at the bottom wall with a higher temperature of $T=1.05T_c$, the temperature at the rest of the bottom and top walls are all fixed at $T_{sat}$ in the simulations.  The physical parameters used in the fluid field are the same as previous cases, while the thermal conductivity here is set to 2/3, and the contact angle at the solid surface is equal to ${75^ \circ }$. In addition, a buoyant force given by ${{\bf{F}}_b} = \left( {\rho  - {\rho _{ave}}} \right){\bf{g}}$ is applied in the vertical direction, in which $ {\rho _{ave}}$ is the mean density over the whole domain, and ${\bf{g}} = \left( {0, - g} \right)$ is the gravity acceleration.

\begin{figure}[H]
	\centering
	\includegraphics[width=0.5\textwidth]{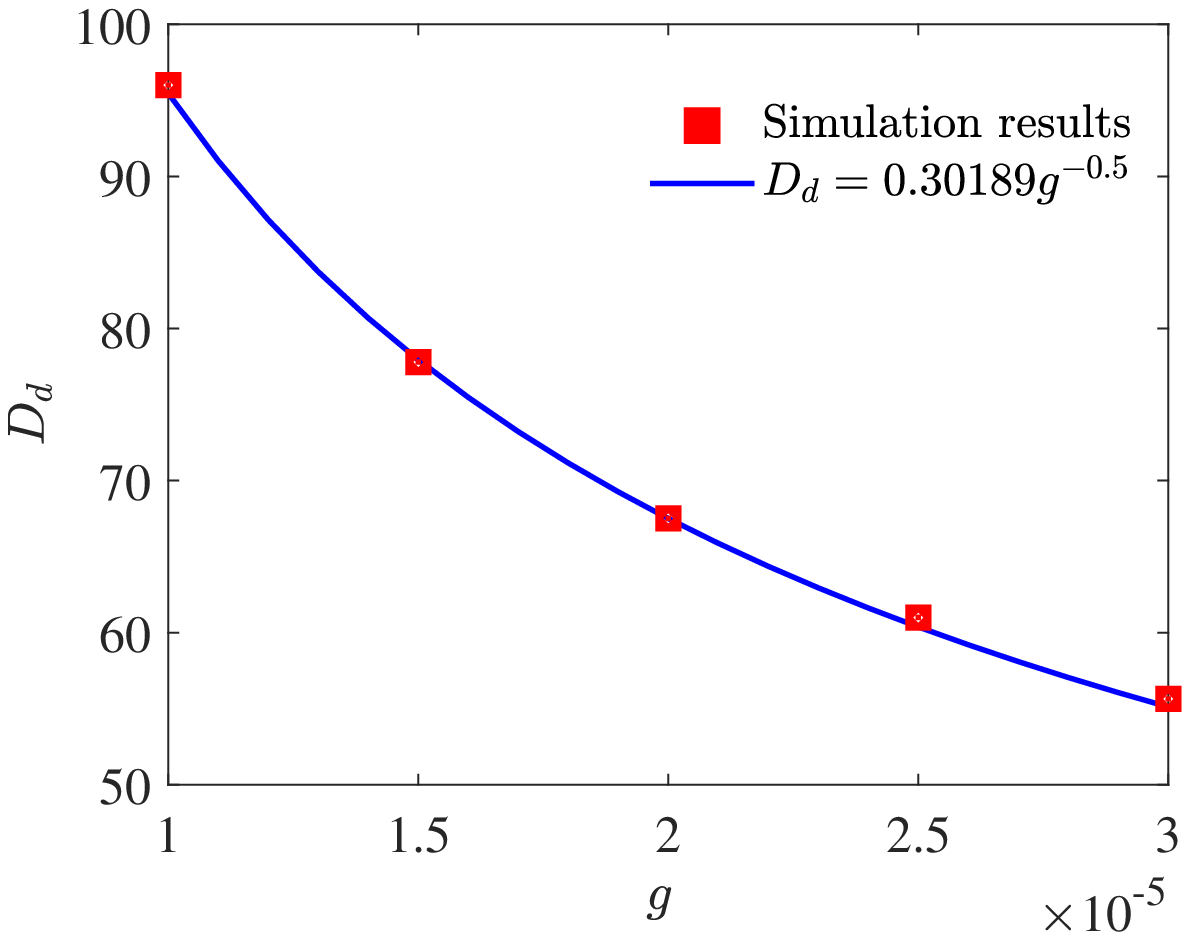}
	\caption{Variation of detachment bubble diameter with gravity acceleration.} 
	\label{fig6}
\end{figure} 

Fig. \ref{fig5} presents several typical snapshots of the nucleate boiling processes. It can be seen from Fig. \ref{fig5} that owing to the influence of the high temperature, a small bubble is initially formed at the center of the solid wall. Then, the size of the bubble gradually increases until it departs from the bottom wall. After that, the detached bubble moves upward under the effect of the buoyancy force, and a new bubble appears at the center of the solid wall, whose behavior is similar to the first bubble. To give a quantitative analysis, we follow the theoretical result given by Fritz \cite{fritz1935maxum}, which states that the relationship between detachment bubble diameter and gravity acceleration satisfying ${D_d} \propto {g^{ - 0.5}}$. To this end, simulations are carried out under different gravity acceleration and the corresponding results are shown in Fig. \ref{fig6}. It is clear that the detachment bubble diameter predicted by the present model is indeed proportional to  $ {g^{ - 0.5}}$, which furthe  illustrates that  our new model is adequate for liquid-vapor phase change problems.

\section{Conclusions}
Liquid-vapor phase change phenomenon often arise in nature and scientific researchers, but numerical modelling of such problem still remains a challenging task in the LB community. The present work proposes a new thermal lattice Boltzmann model for liquid-vapor phase change, which can recover the temperature equation correctly through the multi-scale analysis. In contrast to previous models, the basic idea of the present model is that the temperature equation is treated as a pure diffusion equation with a source term. Additionally, in order to avoid the calculation of the gradient term of $\nabla \left( {\rho {c_v}} \right)$, a novel collision term is  introduced to the evolution equation of the temperature distribution function, which makes the the present model retain the main merits of the LBM.  Several numerical tests show that our new approach could provide comparable results to the WENO scheme, and it is expected to be an efficient method in modeling liquid-vapor phase change.

\end{document}